\documentclass[twocolumn,aps,prl,superscriptaddress,amsmath,amssymb]{revtex4}
\usepackage{amssymb}
\usepackage{amsmath}
\usepackage{amstext}
\usepackage{array}
\usepackage{graphicx}
\usepackage{xspace}
\usepackage{color,soul}

\def\kR{k_{\rm R}}

\def\nuC{\nu_{\rm C}}
\def\nuA{\nu_{\rm A}}
\def\nuR{\nu_{\rm R}}
\def\nuCtL{\tilde{\nu}_{\rm C}^{\rm L}}
\def\nuCtD{\tilde{\nu}_{\rm C}^{\rm D}}
\def\nuAtL{\tilde{\nu}_{\rm A}^{\rm L}}
\def\nuAtD{\tilde{\nu}_{\rm A}^{\rm D}}
\def\nuCL{\nu_{\rm C}^{\rm L}}
\def\nuCD{\nu_{\rm C}^{\rm D}}
\def\nuAL{\nu_{\rm A}^{\rm L}}
\def\nuAD{\nu_{\rm A}^{\rm D}}
\def\nuSC{\nu_{\rm SC}}
\def\nuSA{\nu_{\rm SA}}
\def\phC{\phi_{\rm C}}
\def\phA{\phi_{\rm A}}
\def\phR{\phi_{\rm R}}
\def\phCn{\phi_{\rm C,0}}
\def\phAn{\phi_{\rm A,0}}
\def\phRn{\phi_{\rm  R,0}}
\def\phRmax{\phi_{\rm R, max}}
\def\DphRmax{\Delta \phi_{\rm R,max}}
\def\phQ{\phi_{\rm Q}}
\def\phSC{\phi_{\rm SC}}
\def\phSA{\phi_{\rm SA}}
\def\phSCNull{\phi_{\rm SC}^0}
\def\phSANull{\phi_{\rm SA}^0}
\def\KC{K_{\rm C}}
\def\KA{K_{\rm A}}

\def\JR{J_{\rm R}}
\def\JC{J_{\rm C}}
\def\JA{J_{\rm A}}
\def\CO2{${\rm CO_2}$}
\def\N2{${\rm N_2}$}
\def\O2{${\rm O_2}$}

\def\MR{M_{\rm R}}

\def\chSC{\chi_{\rm SC}}
\def\chSA{\chi_{\rm SA}}

\def\cC{[{\rm C}]}
\def\cN{[{\rm N}]}

\def\phCt{\phi_{\rm C}^*}
\def\phAt{\phi_{\rm A}^*}
\def\phRt{\phi_{\rm R}^*}
\def\phSCt{\phi_{\rm SC}^*}
\def\phSAt{\phi_{\rm SA}^*}

\def\Ta{T_{\rm a}}

\newcommand{\fref}[1]{Fig.~\ref{fig:#1}}

\newcommand{\frefsrange}[2]{Figs.~\ref{fig:#1}-\ref{fig:#2}}
\newcommand{\flabel}[1]{\label{fig:#1}}
\newcommand{\eref}[1]{Eq.~\ref{eqn:#1}}

\newcommand{\erefstwo}[2]{Eqs.~\ref{eqn:#1}~and~\ref{eqn:#2}}

\newcommand{\erefsrange}[2]{Eqs.~\ref{eqn:#1}-\ref{eqn:#2}}
\newcommand{\elabel}[1]{\label{eqn:#1}}

\newcommand{\avg}[1]{\langle #1\rangle}

\begin{document}

\date{\today}

\title{Theory of circadian metabolism}

\author{Michele Monti} \affiliation{FOM Institute AMOLF,
  Science Park 104, 1098 XE Amsterdam, The Netherlands}
\author{David K. Lubensky}\affiliation{Department of Physics,
    University of Michigan, Ann Arbor, MI 48109-1040}
 \author{Pieter
  Rein ten Wolde} \affiliation{FOM Institute AMOLF, Science Park 104,
  1098 XE Amsterdam, The Netherlands}

\begin{abstract}
  Many organisms repartition their proteome in a circadian fashion in
  response to the daily nutrient changes in their environment. A
  striking example is provided by cyanobacteria, which perform
  photosynthesis during the day to fix carbon. These organisms not
  only face the challenge of rewiring their proteome every 12 hours,
  but also the necessity of storing the fixed carbon in the form of
  glycogen to fuel processes during the night. In this
  manuscript, we extend the framework developed by Hwa and coworkers
  (Scott {\it et al.}, Science {\bf 330}, 1099 (2010)) for quantifying
  the relatinship between growth and proteome composition to circadian
  metabolism. We then apply this framework to investigate the
  circadian metabolism of the cyanobacterium {\it Cyanothece}, which
  not only fixes carbon during the day, but also nitrogen during the
  night, storing it in the polymer cyanophycin. Our analysis reveals
  that the need to store carbon and nitrogen tends to generate an
  extreme growth strategy, in which the cells predominantly grow
  during the day, as observed experimentally. This strategy maximizes
  the growth rate over 24 hours, and can be quantitatively understood
  by the bacterial growth laws. Our analysis also shows that the slow
  relaxation of the proteome, arising from the slow growth rate, puts
  a severe constraint on implementing this optimal strategy. Yet, the
  capacity to estimate the time of the day, enabled by the circadian
  clock, makes it possible to anticipate the daily changes in the
  environment and mount a response ahead of time. This significantly
  enhances the growth rate by counteracting the detrimental effects of
  the slow proteome relaxation.
\end{abstract}

\maketitle

\section{Introduction}
Bacterial cells alter gene expression in response to nutrient changes
in their environment
\cite{Scott:2010cx,You:2013ey,Hui:2015ig,Erickson:2017ic,Mori:2017cz}. In
recent years, experiments have demonstrated that the relation between
the proteome composition and the growth rate can be quantitatively
described by growth laws, which are based on the idea that cells need
to balance the supply of amino-acids via catabolic and anabolic
reactions with the demand for amino-acids in the synthesis of proteins
by ribosomes
\cite{Scott:2010cx,You:2013ey,Hui:2015ig,Erickson:2017ic,Mori:2017cz}.
While in the original studies this relationship was tested for
conditions that do not vary on the timescale of the cellular response
\cite{Scott:2010cx,You:2013ey,Hui:2015ig}, more recently it has been
demonstrated that these growth laws can also describe the transient
relaxation dynamics of the proteome in response to a nutrient shift
\cite{Erickson:2017ic,Mori:2017cz}. Here, we extend this framework to
predict how bacterial cells repartition their proteome in response to
periodic, circadian environmental changes.

Many organisms, ranging from cyanobacteria, to plants, insects, and
mammals, possess a circadian clock, which means that they can
anticipate daily changes in their environment, and adjust their
proteome ahead of time. Moreover, many organisms face the challenge
that they can fix carbon and/or nitrogen only during one part of the
day, which means that they then need to store these resources to fuel
processes the other part of the day. In this manuscript, we study by
mathematical modeling the optimal strategy for allocating cellular
resources that maximizes the growth rate of cyanobacterial cells
living in a periodic environment. We show that storing carbon and
nitrogen puts a fundamental constraint on the growth rate, and tends
to generate extreme growth behavior, where cells predominantly grow in
one part of the day. Moreover, we show that in cyanobacteria with
cell-doubling times that are typically longer than 10h
\cite{Sinetova:2012bq,Teng:2013cf,Beliaev:2014jf}, the slow relaxation
of the proteome severely limits the growth rate, but that anticipation
makes it possible to alleviate the detrimental effects of the slow
relaxation.


Cyanobacteria are among the most studied and best characterized
organisms that exhibit circadian metabolism. Their metabolism is
shaped by the constraint that not all the principal elements can be
fixed during the day and the night. For cyanobacteria, the primary
source of carbon is \CO2, which they fix during the day via
photosynthesis. Yet, cyanobacteria also need carbon during the night,
not only for protein synthesis, but also for the generation of fuel
molecules such as ATP, required for maintenance processes such as DNA
repair. To this end, they use not all the fixed carbon to fuel growth
during the day: they also store a fraction in the form of glycogen,
which then becomes the principal source of carbon during the night.

Like all living cells, cyanobacteria not only need carbon, but also
nitrogen. Some cyanobacteria, such as {\it Synechococcus} and {\it
  Synechocystis}, rely on nitrogen that has been fixed by other
organisms in the form of, e.g., nitrate. Other cyanobacteria, such as
{\it Cyanothece} \cite{Reddy:1993te} and {\it Anabaena} 
\cite{Haselkorn:1978uo,Flores:2010cn}, have, however, the ability to fix the
nitrogen that is available in the form of the most abundant gas in the
atmosphere, \N2. Yet, this process requires an enzyme, nitrogenase,
which cannot tolerate \O2. Since \O2 is produced during
photosynthesis, cyanobacterial cells cannot simultaneously fix carbon
and nitrogen during the day. {\it Anabaena} has solved this problem at
the population level, where some cells fix carbon while others fix
nitrogen \cite{Flores:2010cn}. {\it Cyanothece} has solved the problem
at the single-cell level by temporally separating these processes
\cite{Reddy:1993te}. In this manuscript, we will use {\it Cyanothece}
as a model organism to study the design principles of circadian metabolism.

During the day {\it Cyanothece} stores carbon in the form of glycogen,
while during the night it fixes nitrogen and stores it in the form of
cyanophycin \cite{Schneegurt:1994uk}. Like glycogen, cyanophycin is a
large polymer that accumulates in the cytoplasm in the form of
insoluble granules. The polymer is a large polypeptide and consists of
two amino-acids: aspartic acid, which forms the back bone, and
arginine, which constitutes the side group. Arginine is the amino acid
with the largest number of nitrogen atoms in its side chain, namely 3;
indeed, its side chain has the largest ratio of nitrogen (N) to carbon
(C) atoms: 3:4. Cyanophycin is thus exceedingly rich in nitrogen,
having an N:C ratio of about 1:2, which is about an order of magnitude
larger than that in typical proteins. While cyanophycin may serve as a
carbon-storage compound, its principal role is therefore believed to
serve as a nitrogen reservoir.

Under LD conditions, Cyanothece fix nitrogen in the dark, as measured
by the nitrogenase activity, and store glycogen during the day
\cite{Schneegurt:1994uk}. Also in continuous light
\cite{Schneegurt:1994uk} or continuous dark conditions
\cite{Schneegurt:2000vv}, the nitrogenase activity and cyanophycin
storage peak during the subjective night while glycogen storage peaks
during the subjective day, indicating the presence of a circadian
clock that coordinates these activities. Interestingly, under LD
conditions, {\it Cyanothece} exclusively grows during the day
\cite{Cerveny:2009gz}, but in continuous dark, when grown on glycerol
\cite{Schneegurt:2000vv}, it still predominantly grows during the
subjective day, although Fig. 8 of Ref. \cite{Schneegurt:2000vv}
leaves open the possibility it may also grow during the subjective
night.

These physiological rhythms of {\it Cyanothece} are mirrored by
circadian rhythms in gene expression
\cite{Stockel:2008bd,Toepel:2008ez,Stockel:2011ia,Aryal:2011bc,Aryal:2011ho,Welkie:2014ho}. About
30\% of the 5000 genes examined exhibit oscillating expression
profiles \cite{Stockel:2008bd}. Moreover, these genes are primarily
involved in core metabolic processes, such as photosynthesis,
respiration, energy metabolism, and amino-acid biosynthesis
\cite{Stockel:2008bd}; in contrast, most genes involded in transport,
DNA replication and repair, were not differentially expressed
\cite{Stockel:2008bd}. Importantly, genes associated with nitrogen
fixation are primarily expressed in the dark, while those underlying
photosynthesis are upregulated during the ligth and downregulated
during the dark period \cite{Stockel:2008bd}.  Proteomic analysis
using partial metabolic heavy isotope labeling identified 721 proteins
with changing levels of isotope incorporation \cite{Aryal:2011bc}, of
which 425 proteins matched the previously identified cycling
transcripts \cite{Stockel:2008bd}. In particular, the nitrogen
fixation proteins were most abundant during the dark
\cite{Aryal:2011bc} while many proteins involved in photo-synthesis
are present in higher abundance during the light. Interestingly,
proteins involved in storing glycogen, such as the glycogen synthase,
peak during the light, while enzymes involved in glycogen metabolism,
such as glycogen phosphorylase, GlgP1, have higher levels during the
dark \cite{Stockel:2008bd}. Conversely, the cyanophycin processing
enzyme cyanophycinase, CphB, which breaks down cyanophycin into
arginine and aspartic acid, shows higher synthesis in the light
\cite{Aryal:2011bc}, although, perhaps surprisingly, cyanophycin
synthetase, dCphA, appears not to be strongly coupled with the
light-dark cycle.

These transcriptomic and proteomic analyses
\cite{Stockel:2008bd,Toepel:2008ez,Stockel:2011ia,Aryal:2011bc,Welkie:2014ho},
together with large-scale computational modeling of the metabolic
network \cite{Reimers:2017dq}, provide detailed information about the
proteome repartitioning dynamics during the 24 h period. Yet, many
questions remain open: First and foremost, why do cyanobacterial cells
typically exclusively grow during the day? Cyanobacterial cells have
the components to grow at night, which suggests that the strategy to
not grow during the dark arises from a cellular trade-off that
maximizes the growth rate over 24h \cite{Reimers:2017dq}. Can this
trade-off be quantified, and do cellular growth laws predict that it
is optimal to not grow at all during the night?  Secondly, in the
absence of active protein degradation, the timescale for the
relaxation of the proteome is given by the growth rate
\cite{ThesisVandenEnde,Erickson:2017ic,Mori:2017cz}, while at the same
time the growth rate of these cyanobacterial cells is affected by how
fast the proteome can adjust to the changing light and nutrient levels
(glycogen and cyanophycin). This observation is particularly
pertinent, because the growth rate of these cyanobacterial cells tend
to be low, with cell-division times that are typically longer than 10
hours \cite{Sinetova:2012bq,Teng:2013cf,Beliaev:2014jf}. How much is
the growth rate limited by the slow relaxation of the proteome?
Thirdly, cyanbacterial cells have a circadian clock, which allows them
to predict and anticipate the changes in light and nutrient levels. In
general, anticipation becomes potentially beneficial especially when
the cellular response is slow \cite{Becker:2015iu}. Does anticipation
allow cyanobacterial cells to significantly raise their growth rate?

To address these questions, we employ the framework developed by Hwa
and coworkers for quantifying the relationship between growth and
proteome composition
\cite{Scott:2010cx,You:2013ey,Hui:2015ig,Erickson:2017ic,Mori:2017cz}
and extend it to describe circadian metabolism. This framework is
inspired by two key observations: On the one hand the response to a
changing environment tends to be extremely complex at the molecular
level, involving a myriad of signaling and metabolic pathways. On the
other hand, it tends to be global, meaning that in response to a
nutrient limitation certain subsets of enzymes are upregulated while
others are downregulated. The system is therefore not described in
terms of the detailed signaling and metabolic networks, but rather via
coarse-grained protein sectors. Each sector contains a subset of
enzymes, which share a common purpose, according to the
supply-and-demand picture of protein synthesis \cite{Hui:2015ig}. 
Each sector is described by a single coarse-grained enzyme, which can
be thought of as representing the average activity of the enzymes in
that sector. It is this coarse-grained description that allows for a
quantitative mathematical analysis.  The framework has been used to
describe the effect of protein overexpression \cite{Scott:2010cx},
cAMP-mediated catabolite repression \cite{You:2013ey}, growth
bistability in response to anti-biotics \cite{Deris:2013ia}, and
methionine biosynthesis \cite{Li:2014ba}. And importantly for our
analysis, it has recently been extended to describe the transient
relaxation dynamics of the proteome in response to a nutrient shift
\cite{Erickson:2017ic,Mori:2017cz}. While these studies have focused
on the bacterium {\it Escherichia coli}, we here employ this framework
to study circadian metabolism of cyanobacteria.

The model that we present aims to describe the circadian metabolism of
cyanobacteria like {\it Cyanothece}, which fix carbon during the day
and nitrogen during the night, although it can straightforwardly be
amended to describe the metabolism of cyanobacteria such as {\it
  Synechococcus} and {\it Synechocystis} that only fix
carbon. Arguably the most minimal model to capture the interplay
between carbon and nitrogen fixation is one that consists of a
ribosome sector, a carbon sector and a nitrogen
sector. However, to capture the fact that storing glycogen and
cyanophycin does not directly contribute to growth, but only
indirectly, by providing carbon and nitrogen the next part of the day,
our model also contains two other protein sectors: a glycogen and a
cyanophycin synthesis sector. Our model therefore naturally includes
two important consequences of building a carbon and nitrogen
reservoir: 1) it requires the synthesis of enzymes that do not
directly contribute to growth, and hence lower the instantaneous
growth rate \cite{Scott:2010cx}; 2) storing carbon and nitrogen atoms
drains carbon and nitrogen flux away from
protein synthesis. Our model further incorporates the dynamics of the
carbon and nitrogen reservoirs (glycogen and cyanophycin), the slow
relaxation of the proteome in response to the changing nutrient levels,
and the capacity to anticipate
the changing nutrient levels by mounting a response ahead of time.

We first use this model to study the optimal strategy that maximizes
the growth rate over 24 hours. Our analysis reveals that the need to
store carbon and nitrogen tends to generate an extreme strategy, in
which cells predominantly grow during the day, as observed
experimentally \cite{Schneegurt:2000vv,Cerveny:2009gz}. However, our
analysis also reveals that the slow relaxaton of the proteome, arising
from the slow growth rate, puts a severe constraint on implementing
this optimal strategy. In essence, to store enough cyanophcyin during
the night to fuel growth during the day, the cyanophycin-storing
enzymes need to be expressed at levels that cannot be reached if the
cells would only start expressing these enzymes at night. Indeed, to
implement the optimal strategy, the cells need to express these
enzymes already before the beginning of the night, when they still
grow significantly. Interestingly, recent transcriptomics and
proteomics data provide evidence for this prediction
\cite{Welkie:2014ho}.



\section{Theory} 
The central ingredients of the framework of Hwa and coworkers
\cite{Scott:2010cx,You:2013ey,Hui:2015ig,Erickson:2017ic,Mori:2017cz}
are the coarse-grained protein sectors and the balance of fluxes
between them. We describe these elements in turn.

{\bf Protein setors} The sectors are defined experimentally by how the
enzyme expression levels vary with the growth rate in response to
different types of nutrient limitation \cite{Hui:2015ig}. The C-sector
is defined as the subset of enzymes whose expression levels increase
as the growth rate decreases upon a Carbon limitation, yet decrease as
the growth rate decreases upon a nitrogen limitation or translation
inhibition \cite{Hui:2015ig}. A mass-spec analysis of {\it E. coli}
revealed that this sector contains enzymes involved in ion-transport,
the TCA- cycle and locomotion \cite{Hui:2015ig}. The A-sector is
defined as the group of proteins that are upregulated in response an
A-limitation---a nitrogen limitation--- yet downregulated in response
to carbon or translation limitation. In {\it E. coli}, this sector
consists of enzymes that are involved in the incorporation of nitrogen
into amino-acids \cite{Hui:2015ig}. The R-sector contains the
ribosomes, which increase in abundance as the growth rate decreases
upon the addition of a translation inhibitor. The study of Hui and
coworkers on {\it E. coli} also identified an S-sector, consisting of
enzymes whose expression levels increase in response to both carbon
and nitrogen limitation, and a U-sector, consisting of proteins that
are uninduced under any of the applied limitations \cite{Hui:2015ig}.

In our model, we are interested in the interplay between carbon and
nitrogen assimilation, and the simplest model that can capture this
interplay is one that considers an R-sector, a
C-sector and an A-sector. The mass fractions of the proteins in these sectors
are denoted by $\phR$, $\phC$ and $\phA$, respectively. Our model does
not explicly contain an S- and a U-sector, although we emphasize that
as experimental data becomes available the model can straightfowrardly
be extended to include these sectors \cite{Hui:2015ig}. Following Hui
{\it et al.}, we also stress that these sectors are ultimately defined
experimentally \cite{Hui:2015ig}. In our case, the C-sector is defined as consisting of
those enzymes that are upregulated in response to a carbon limitation,
yet downregulated in response to an A- or R-limitation. The carbon
limitation can be in the form of reduced \CO2 and light levels during
the day, but also reduced glycogen levels during the night. Our model
thus lumps all proteins that are involved in providing carbon
skeletons for amino-acid synthesis into one sector, the C-sector. We
anticipate that this sector contains enzymes of not only the
photosynthesis machinery, but also the TCA cycle, as well as enzymes
involved in degrading glycogen, such as glycogen phosphorylase
GlgP. Experiments need to establish whether it would be necessary to
split this C-sector up into separate sectors for, e.g.,
photosynthesis, glycogen breakdown and downstream carbon
processing (e.g. TCA cycle). 

Similarly, we define the A-sector as the set of enzymes that are
upregulated in response to nitrogen limitation, yet downregulated in
resopnse to a C- or R-limitation. We envision that nitrogen limitation
can be imposed by reducing \N2 levels, by employing a titratable
nitrogen uptake system \cite{Hui:2015ig}, or by lowering levels of
cyanophycin. While, again, experiments need to identify which enzymes
belong to this sector, we expect that it contains not only the
nitrogenase enzymes that reduce \N2 into ammonia and the enzymes that
subsequently incorporate the nitrogen into amino-acids, but also the
proteins that are involved in the breakdown of cyanophycin, such as
cyanophycinase CphB.

Following Scott {\it et al.}, the model also
includes an unresponse fraction $\phQ$, although this parameter will
be absorbed in the maximal ribosomal fraction $\phRmax$, as described
below \cite{Scott:2010cx,You:2013ey,Hui:2015ig}.

{\bf Storing fractions} While this model is indeed highly
coarse-grained, we do consider two other sectors, which contain
enzymes that store glycogen during the day and nitrogen during the
night. Their fractions are denoted by $\phSC$ and $\phSA$,
respectively. Enzymes belonging to $\phSC$ are glycogenin and glycogen
synthase, which are indeed upregulated during the day
\cite{Stockel:2011ia,Aryal:2011bc}. Cyanophycin is synthesized from
arginine and aspartate by a single enzyme, cyanophyinc synthetase,
CphA, which thus forms the $\phSA$ sector \cite{Stockel:2011ia}. A key
point is that expressing the glycogen-storing enzymes slows down
growth during the day yet enables growth during the night, while
expressing cyanophycin synthetase sloww down growth during the night,
yet enables growth during the day. Even though the growth laws are
linear, this creates a feedback between growth at night and
during the day that yields a non-linear response, as we discuss in
more detail below.

{\bf Steady-state flux balance}
The experiments by Hwa and coworkers on {\it E. coli} have revealed
that the steady-state growth rate varies linearly with the size of the protein
sectors \cite{Scott:2010cx,You:2013ey,Hui:2015ig}. These linear
relationships can be understood by combining the following ideas: a)
in steady-state the fluxes $J_\alpha$ through the different sectors
$\alpha={\rm R, C, A}$ are balanced, so that there is no build-up of
intermediates like amino-acids; b) the growth rate $\lambda$ is
proportional to the flux through the sectors; c) the flux through a
sector scales linearly with the size of the sector. Combining these
ideas makes it possible to quantitatively describe the experiments of
\cite{You:2013ey,Hui:2015ig}, explaining how each sector
is upregulated in response to one type of limitation, while
downregulated in response to another type of limitation
\cite{You:2013ey,Hui:2015ig}. Moreover, the model can
quantitatively describe how the growth rate decreases as an unnecessary
protein, which does not directly contribute to growth, is expressed
via an artificial inducer \cite{Scott:2010cx}. The latter is
important, because the proteins that store glycogen and
cyanophycin, respectively, can be thought of as proteins that do not
directly contribute to growth; they only contribute by providing the
carbon- and nitrogen sources for the next part of the day. Our model
incorporates these three ingredients a)-c), but adds a fourth, d): a
certain fraction of the flux through the carbon and nitrogen sector is
reserved for storing glycogen during the day and cyanophycin during
the night, respectively.

While our full model is time dependent, we will first consider a
simpler model in which we can directly use the growth laws derived by
Hwa and coworkers
\cite{Scott:2010cx,You:2013ey,Hui:2015ig}. Specifically, during the
day the principal source of carbon is \CO2, while that of nitrogen is
cyanophycin, which decreases with time. During the night, the
principal source of nitrogen is \N2, while that of carbon is glycogen,
which falls with time. As a result, the growth rate $\lambda$ will, in
general, be time-dependent, $\lambda=\lambda(t)$. Moreover, because
the glycogen and cyanophycin concentrations vary with time, the
proteome fractions, which are adjusted in accordance with the nutrient
availability, will change not only upon the shift from day to night,
but also continue to change throughout the day and night. As was pointed
out in \cite{ThesisVandenEnde} and also in
\cite{Mori:2017cz,Erickson:2017ic}, in the absence of active protein
degradation, the proteome relaxes with a timescale that is set by the
growth rate $\lambda(t)$. This means that when the growth rate is low,
the proteome will relax slowly, and may not be in quasi-equilibrium
with respect to the instantaneous levels of glycogen during the night
and cyanophycin during the day. Below we will take this
slow relaxation of the proteome into account. However, to introduce
the main elements of the model, it will be instructive to first assume
that the growth rate is so high, that the proteome is always in
quasi-equilbrium with respect to the instantaneous nutrient levels,
set by the \CO2/light and cyanophycin levels during the day, and the
glycogen and \N2 levels during the night. The growth thus depends on
time, but not explicitly, and only implicitly via the levels of
glycogen and cyanophycin: $\lambda(t) = \lambda([{\rm C}](t),[{\rm
  N}](t))$, where $\cC(t)$ and $\cN(t)$ are the time-dependent carbon
and nitrogen sources. We call this model the quasi-equilibrium model.

{\bf Quasi-equilibrium model} The first two ingredients a) and b)
imply that in the quasi-equilibrium model
\begin{align}
\lambda^\beta(t) = c_{\rm R} \JR (t)= c_{\rm C} \JC^\beta (t) = c_{\rm
  A} \JA^\beta (t),
\elabel{FB}
\end{align}
where $c_{\alpha}$ are the stochiometric requirements for cell growth
\cite{Hui:2015ig}. Here, we have added the superscript $\beta={\rm L,D}$, with ${\rm
  L}$ standing for light and ${\rm D}$ for dark, to remind ourselves
that the fluxes through the carbon and nitrogen sector and thereby the
growth rate, depend on the source of carbon and nitrogen used, which
differs between day and night. 

The third observation c) means that for the ribosomal sector
\begin{align}
\lambda^{\rm L/D}(t) &=c_{\rm R} \JR (t) = \nuR (\phR (t) - \phRn).
\elabel{JR}
\end{align}
Here, $\nuR = c_{\rm R} \kR$, where $\kR$ describes the translation
efficiency, which can be varied experimentally using a translation
inhibitor such as cloramphenicol
\cite{Scott:2010cx,You:2013ey,Hui:2015ig}. The quantity $\phRn$ is the
fraction of ribosomes that is not active in steady-state, yet can
become active during the transition from one environment to the next
\cite{Erickson:2017ic,Mori:2017cz}. In the quasi-equilibrium model
considered here, it is a constant, independent of time. 

The ingredients c) and d) imply that the flux through the carbon
sector that flows into the other sectors is given by
\begin{align} 
\lambda^{\beta} (t)&=  c_{\rm C}\JC^\beta (t) = \nuC^\beta (t) (\phC (t) -\phCn) - L(t) \nuSC \phSC (t).
\elabel{JC}
\end{align}
Here, $L(t)$ is an indicator function that is 1 during the day and 0
during the night. Indeed, during the day both terms are present. The
first term on the right-hand side describes the carbon flux that would
flow into the other sectors if no carbon were stored into glycogen
during the day. The second term indeed describes the flux that is not
used for growth during the day, but rather lost in storing
glycogen. During the night, no glycogen is stored and the second term
is absent.  In the first term, $\nuC^{\beta}$ is a measure for the
efficiency of the carbon sector. It depends on the quality and the
amount of nutrient \cite{You:2013ey,Hui:2015ig}, but can also be
varied experimentally---in {\it E. coli} by titrating a key enzyme
such as the lactose permease \cite{You:2013ey,Hui:2015ig}. In our
model, $\nuC^\beta$ depends on the part of the day, as indicated by
$\beta = {\rm L,D}$: during the day, the principal carbon source is
\CO2, which means that the value of $\nuCL$ will depend on the
concentration of \CO2 and light levels. Since we will model the light
intensity as a step function, during the day the light level and hence
$\nuCL(t)$ is constant, and equal to $\nuCL(t)=\nuCtL$.  In contrast,
during the night, the principal source of carbon is glycogen, which
decreases during the night. This affects the carbon-processing
efficiency. We will model this as
\begin{align}
\nuCD(t) = \nuCtD \frac{[{\rm C}](t)}{[{\rm C}](t)+\KC},
\elabel{nuCD}
\end{align}
where $[{\rm C}](t)$ is the time-dependent concentration of glycogen and
$\KC$ is the glycogen concetration at which the enzyme efficiency is
reduced by a factor of 2. The quantiy $\nuCtD$ is the maximal
efficiency of the carbon-sector with glycogen as the carbon source; it
does not depend on time. The quantity $\phCn$ is the fraction of
carbon-processing enzymes that is not used for growth. For {\it
  E. coli} it is very close to zero, and from hereon we assume it to
be zero. The quantity $\nuSC$ describes the efficiency of the
glycogen-storing enzymes, and is taken to be constant.

For the nitrogen-sector, we similarly obtain
\begin{align}
\lambda^\beta(t)&=c_{\rm A} \JA^\beta (t)\nonumber\\
& = \nuA^\beta (t) (\phA(t) - \phAn) - (1-L(t)) \nuSA\phSA (t),
\end{align}
where in the calculations performed here we assume that $\phAn$ is
zero, even though the experiments indicate that for {\it E. coli} the
unused fraction in the A-sector is about 10\% \cite{Hui:2015ig}. The
nitrogen-processing efficiency during the day depends on the concentration of
stored cyanophycin, $[{\rm N}]$, via
\begin{align}
\nuAL (t) = \nuAtL \frac{[{\rm N}](t)}{[{\rm N}](t)+\KA}.
\elabel{nuAL}
\end{align}
The nitrogen-uptake efficiency during the night depends on the amount
of \N2, which we assume to be constant throughout the night. The
efficiency is thus given by $\nuAD=\nuAtD$.

Combining all four ingredients a) - d), i.e. \erefsrange{FB}{nuAL}, yields
\begin{align}
  \lambda^\beta(t) &=\nuR (\phR (t) - \phRn) \elabel{lambda_kR}\\
   & = \nuC^\beta (t) \phC(t) - L(t) \nuSC \phSC(t) \elabel{lambda_kC}\\
  &=\nuA^\beta (t) \phA - (1-L(t)) \nuSA \phSA (t) \elabel{lambda_kA}.
\end{align}

{\bf Proteome balance}
The protein sectors obey at all times $t$ the constraint
\begin{align}
\phR(t) + \phC(t) + \phA(t) + \phSC(t) + \phSA(t) + \phQ = 1.
\end{align}
The growth rate $\lambda$ will be maximal, $\lambda\to\lambda_{\rm max}$, when
the storing, carbon- and nitrogen-processing fractions approach zero, and the
ribosomal fraction becomes maximal
\begin{align}
\lim_{\lambda \to \lambda_{\rm max}} \phR&\equiv \phRmax = 1- \phQ,
\end{align}
allowing us to rewrite the constraint as:
\begin{align}
\phR(t) + \phC(t) + \phA(t) + \phSC(t) + \phSA(t) = \phRmax.
\elabel{constr_phiRmax}
\end{align}
We note that this definition of $\phRmax$ differs slightly from $\phi_{\rm
  max}$ defined in Ref. \cite{You:2013ey,Hui:2015ig}.

{\bf Growth rate in quasi-equilibrium model} In our model, the input
parameters are $\nu_\alpha^\beta$ and $\phRn$, while the storing
fractions $\phSC(t)$ and $\phSA(t)$ are control parameters over which
we will optimize to maximize the growth rate over a 24h period. In
the quasi-equilibrium model, the optimal $\phSC$ during the night is
zero and the optimal $\phSA$ during the day is zero. In this model, we
thus have one optimization parameter $\phSC$ for the day, and another
for the night, $\phSA$. In this quasi-equilibrium model, the other
protein sectors relax instantaneously, to values that, for the day,
are determined by the efficiencies $\nuR$, $\nuCL$, the instantaneous
efficieny $\nuAD(t)$ and the optimization parameter $\phSC(t)$, and, for
the night to values given by $\nuR$, $\nuAD$,  the instantaneous value of
$\nuAL(t)$ and the optimization parameter $\phSA(t)$.
The 4 equations,
\erefsrange{lambda_kR}{lambda_kA} together with the constraint
\eref{constr_phiRmax}, thus contains 4 unknowns $\phR, \phC, \phA,
\lambda$, which can be solved to obtain the instantaneous growth rate
for the day and night, respectively, for the quasi-equilibrium model:
\begin{align}
\lambda^{\rm L}(t) &= \frac{\nuR \nuCL \nuAL(t)}{\nuR \nuCL + \nuR \nuAL(t) +
  \nuAL(t) \nuC} \times \nonumber \\
&(\phRmax - \phRn - (1+\nuSC/\nuCL) \phSC(t)) \elabel{lambdaL}\\
\lambda^{\rm D}(t) &=\frac{\nuR \nuCD(t) \nuAD}{\nuR \nuCD(t) + \nuR \nuAD +
  \nuAD \nuCD(t)}\times\nonumber\\
&(\phRmax - \phRn - (1+\nuSA/\nuAD) \phSA(t)) \elabel{lambdaD}
\end{align}
Clearly, during the day the growth rate, for given \CO2 and
cyanophycin levels, is maximal when
no glycogen is stored and $\phSC=0$. This defines a maximum growth
rate during the day
\begin{align}
\lambda_{\rm max}^{\rm L} ([{\rm N}](t)) &=\frac{\nuR \nuCL \nuAL(t)}{\nuR \nuCL + \nuR \nuAL(t) +
  \nuAL(t) \nuCL} \Delta \phRmax,\elabel{lambdamaxL}
\end{align}
where $\DphRmax = \phRmax - \phRn$.
The maximal growth rate depends on the instantaneous amount of
cyanophycin, $[{\rm N}](t)$, because $\nuAL(t)$ depends on $[{\rm N}](t)$ (see \eref{nuAL}).
From \eref{lambdaL} we find the storing fraction $\phSCNull$
that reduces the growth rate to zero during the day:
\begin{align}
\phSCNull=\frac{\DphRmax}{1+\nuSC/\nuCL}.
\elabel{phSCNull}
\end{align}
 This allows us to rewrite
\eref{lambdaL} as
\begin{align}
\lambda^{\rm L} (\phSC) &= \lambda_{\rm max}^{\rm L}([{\rm N}](t)) \left (1
  - \phSC / \phSCNull\right).
\elabel{lambdaLrw}
\end{align}
Equivalently, we find for the growth rate during the night
\begin{align}
\lambda^{\rm D} (\phSA) &= \lambda_{\rm max}^{\rm D}([{\rm C}](t)) \left (1
  - \phSA / \phSANull\right),
\elabel{lambdaDrw}
\end{align}
with
\begin{align}
\lambda_{\rm max}^{\rm D}&=\frac{\nuR \nuCD(t) \nuAD}{\nuR \nuCD(t) + \nuR \nuAD +
  \nuAD \nuCD(t)}\DphRmax \elabel{lambdamaxD}
\end{align}
 and
\begin{align}
\phSANull=\frac{\DphRmax}{1+\nuSA/\nuAD}.
\elabel{phSANull}
\end{align}

  A few points are worthy of note. Firstly,
  \erefstwo{lambdaLrw}{lambdaDrw} show that the growth rate
  decreases linearly with $\phSC$ and $\phSA$, respectively.  In fact,
  Scott {\it et al.} derived a similar relation for the growth rate
  when an unnecessary protein is expressed \cite{Scott:2010cx}. This
  highlights the idea that storing glycogen and cyanophycin reduces
  the growth rate, because synthesizing these storage molecules
  requires proteins that do not directly contribute to growth---thus
  taking up resources that could have been devoted to making more
  ribosomes. Indeed, building carbon and nitrogen reservoirs only pays
  off the next part of the day, which can be seen by noting that the
  maximum growth rate during the day, $\lambda_{\rm max}^{\rm L}$,
  increases with the amount of cyanophycin that has been stored the
  night before (via $\nuAL$, see \eref{nuAL}), while the maximum growth rate during the
  night, $\lambda_{\rm max}^{\rm D}$ increases with the amount of
  glycogen that has been stored the day before (via $\nuCD$, see \eref{nuCD}). Clearly,
  the cell needs to strike a balance between maximizing the
  instantaneous growth rate and storing enough resources to fuel
  growth the next part of the day. 

  However, there is also another effect: building the reservoirs
  reduces the growth rate not only because it requires proteins that
  do not directly contribute to growth, but also because it drains
  carbon and nitrogen flux. This manifests itself in the intercepts
  $\phSCNull$ and $\phSANull$ at which the growth rate is zero (see
  \eref{phSCNull}). This effect puts a hard fudnamental bound on the
  maximum rate at which glycogen and cyanophycin can be stored. For
  glycogen the maximum storing rate is given by
\begin{align}
v_{\rm store,G}^{\rm max} &= c_{\rm G} \nuSC \phSCNull \elabel{vmax1}\\
&=c_{\rm G} \frac{\nuSC \nuCL}{\nuSC+\nuCL}\DphRmax.\elabel{vmax2}
\end{align}
Here, $c_{\rm G}$ is a stochiometric coefficient that reflects the
number of carbon atoms that are stored in a glycogen molecule. This
expression shows that the maximal storing rate $v_{\rm store,G}^{\rm
  max}$ increases with $\DphRmax$. This is because $\DphRmax$ limits
the fraction of the proteome that can be allocated to storing
glycogen, $\phSCNull$. The expression also reveals that $v_{\rm
  store,G}^{\rm max}$ depends on $\nuSC$ and $\nuCL$. The maximum
storing rate $v_{\rm store,G}^{\rm max}$ initially increases with
$\nuSC$, simply because that increases the rate at which the glycogen
storing enzymes operate. However, the increased flux of carbon into
glycogen also means that less carbon is available for making the
glycogen-storing enzymes themselves. As a result, as $\nuSC$
increases, the maximal fraction $\phSCNull$ of glycogen-storing
enzymes decreases (see \eref{phSCNull}). In the limit that $\nuSC$
becomes very large, i.e. much larger than $\nuC$, then $\phSCNull$
becomes zero, and the rate at which glycogen is stored becomes
independent of $\nuSC$. In this regime, all the carbon flows into
glycogen and the storing rate instead becomes limited by $\nuCL$,
$v_{\rm storge, G}^{\rm max}=c_{\rm G}\nuCL \DphRmax$. In this limit,
$\phR=\phA=0$ and $\phC=\DphRmax$, such that there is no carbon flow
devoted to growth, $\nuCL\phC - \nuSC \phSC =0$ (\eref{JC}), but only
to storing glycogen. As we will see below, this will put a strong
constraint on the maximal growth rate of the cyanobacteria.

{\bf Reservoir dynamics}
The growth rate depends on the efficiencies $\nuCD(t)$ and $\nuAL(t)$,
which depend on the amount of glycogen and cyanophycin, respectively
(see \erefstwo{nuCD}{nuAL}). The dynamics of their
concentration is given by
\begin{align}
\frac{d[{\rm N}](t)}{dt}&=(1-L(t)) c_{\rm CP} \nuSA \phSA(t) - L(t) c_{\rm
  CP} \lambda (t)
- \lambda (t)[{\rm N}](t) \elabel{dNdt}\\
\frac{d[{\rm C}](t)}{[dt]}&=L(t) c_{\rm G} \nuSC \phSC(t) - (1-L(t))c_{\rm G} \lambda (t)
  - \lambda (t) [{\rm C}](t) \elabel{dCdt}
\end{align}
The last term in both equations is a dilution term, where we have
exploited that cells grow exponentially with rate $\lambda(t)$. The
first term describes the accumulation of the stores due to the storing
enzymes, with $c_{\rm CP}, c_{\rm G}$ being stoichiometric coefficients
that reflect how many nitrogen and carbon atoms are stored in a
cyanophycin and glycogen molecule, respectively. The second term
describes the consumption of cyanophycin and glycogen that fuels
growth. Focusing on glycogen, this term can be understood by noting
that the depletion of glycogen during the night is given by the rate
at which the carbon sector consumes glycogen:
\begin{align}
\frac{d[{\rm C}](t)}{dt}&=(1-L(t)) c_{\rm G} \nuCtD 
\frac{\cC(t)}{\cC(t)+\KC} \phC(t)\\
&=(1-L(t)) c_{\rm G} \lambda(t)
\end{align}
where in the second line we have exploited that in quasi-equilibrium
the growth rate $\lambda(t)$ is given by the flux through the carbon
sector (see \erefstwo{JC}{nuCD}). Importantly, this expression reveals
that the depletion of the store depends on the growth rate not only
because that sets the dilution rate (reflected by the third term in
\erefstwo{dNdt}{dCdt}), but also because the growth rate sets the rate
at which the store is consumed (second term).

{\bf Slow proteome dynamics} The proteome will in general not be in
quasi-equilibrium with respect to the instantaneous nutrient
levels. To take into account the relaxation of the proteome we first
define the mass fractions $\phi_\alpha$ of the different sectors
\begin{align}
\phi_\alpha = \frac{M_\alpha}{M}
\end{align}
where $M_\alpha$ is the protein mass of sector $\alpha = {\rm R,C,A, SC,SA,Q}$
and $M$ is the total mass. The total rate at which proteins are synthesized
is given by
\begin{align}
\frac{dM(t)}{dt} &= \sigma(t) \MR (t),
\end{align}
where $\MR$ is the mass of the ribosomal sector, consisting of the
mass of the ribosomes and the ribosome-affiliated proteins
\cite{Mori:2017cz}. The quantity $\sigma(t)$ is the instantaneous
translational efficiency. It corresponds to the average translational
efficiency, and does not distinguish between active and inactive
ribosomes \cite{Mori:2017cz}. When we divide the above equation by $M(t)$, we obtain the
instantaneous growth rate \cite{ThesisVandenEnde,Mori:2017cz}
\begin{align}
\lambda(t) &= \frac{1}{M(t)}\frac{dM(t)}{dt} = \sigma(t) \phi_R(t).
\elabel{ilt}
\end{align}

To obtain the evolution of the different protein sectors, we denote
the fraction of the number of ribosomes that are allocated to making
protein sector $\alpha$ as $\chi_\alpha$.  The evolution
of the proteome mass $M_\alpha$ is then
\begin{align}
\frac{dM_\alpha(t)}{dt} = \chi_\alpha(t) \sigma(t)\MR 
\end{align}
and that of the proteome fraction \cite{ThesisVandenEnde,Mori:2017cz}
\begin{align}
\frac{d\phi_\alpha(t)}{dt} &=\frac{1}{M(t)}\frac{dM_\alpha(t)}{dt} -
\frac{M_\alpha(t)}{M^2(t)} \frac{dM(t)}{dt}\\
&=\frac{1}{M(t)}\chi_\alpha(t) \sigma(t)\MR - \phi_\alpha (t)
\lambda(t)\\
&= \chi_\alpha(t) \sigma(t) \phR(t) - \phi_\alpha(t) \lambda(t)\\
&=\lambda(t) (\chi_\alpha (t) - \phi_\alpha(t)),\elabel{dphadt}
\end{align}
where in going to the last line we have exploited \eref{ilt}. This
equations shows that when $\chi_\alpha(t)$ adjust rapidly to a new
nutrient environment, as recent experiments indicate
\cite{Erickson:2017ic,Mori:2017cz},  the relaxation of the
proteome is dominated by the growth rate $\lambda(t)$. Importantly,
the equation also shows that when $\chi_\alpha(t) = \phi_\alpha(t)$,
the proteome has equilibrated: the fractions no longer change with
time. 

Recent experiments indicate that after a nutrient upshift the
translational efficiency $\sigma(t)$ and the fraction $\chi_\alpha(t)$
of ribosomes devoted to making proteins of sector $\alpha$ rapidly
approach their new steady-state values as set by the new environment
\cite{Erickson:2017ic,Mori:2017cz}. We therefore make the
simplification \cite{ThesisVandenEnde}, also used in
\cite{Mori:2017cz}, that after a day-night (and night-day) transition
$\sigma(t)$ immediately takes the final value $\sigma^*$ set by the
new environment and that $\chi_\alpha(t)$ immediately takes the value
of the final fraction $\phi_\alpha^*$ in the new environment. However,
in our system, the amounts of glycogen and cyanophycin change with
time, and the proteome fractions continually adjust to this. The
``final'' fractions $\phi_\alpha^*$ are thus target fractions that
themselves change with time, and similarly for the translation
efficiency $\sigma$:
\begin{align}
\chi_\alpha(t) &= \phi^*_\alpha(\cC(t),\cN(t))\elabel{chpht}\\
\sigma(t) &=\sigma^*(\cC(t),\cN(t)).\elabel{sigmat} 
\end{align}
These quantities are set such that if $\phi_\alpha(t)$ were equal to
$\chi_\alpha(t) = \phi^*_\alpha(\cC(t),\cN(t))$ and $\sigma(t)$ were equal to
$\sigma^*(\cC(t),\cN(t))$, the fluxes through the different sectors
would be balanced and the growth rate would be equal to $\lambda^*(t)$:
\begin{align}
\lambda^*(t) &=\sigma^*(t) \phRt(t) = \nu_R (\phRt(t) - \phRn) \elabel{JRt}\\
&=\nuC^\beta(\cC(t)) \phCt(t) - L(t) \nuSC \phSCt (t)\elabel{JCt}\\
&=\nuA^\beta(\cN(t)) \phAt(t) - (1-L(t)) \nuSA \phSAt(t)\elabel{JAt}
\end{align}
Importantly, we do not only need to consider the target fractions for
the R-, C-, A-, and Q-sector, but also for the storing fractions:
$\chSC(t)=\phSCt(t)$ and $\chSA(t) = \phSAt(t)$. \erefsrange{JRt}{JAt}
are thus solved subject to the following constraint
\begin{align}
\phRt(t) + \phCt(t) + \phAt(t) + \phSCt(t) + \phSAt(t) = \phRmax,\elabel{TargetConstraint}
\end{align}
where $\phSC(t)$ and $\phSA(t)$ are optimization parameters described
in more detail below.  This equation states that the total ribosome
protein synthesis fraction $\sum_\alpha \chi_\alpha(t)=1$ at all
times, which guarantees that $\sum_\alpha \phi_\alpha(t)=1$ at all
times.

{\bf Anticipation} The cells need to repartition its proteome every 12h as the cells move from day to night, and vice versa. Moreover,
the cells need to continually adjust its proteome to the changing
levels of cyanophycin and glycogen. However, the relaxation of the
proteome is, in the absence of protein degradation, set by the growth
rate (see \eref{dphadt}), which for cyanobacteria, with cell division times
in the range of 10 - 70h, is low compared to the 24 hr period of
the day-night cycle. This slow relaxation of the proteome will tend to
make the growth rate suboptimal. Interestingly, cyanobacteria, ranging
from {\it Synechococcus}, {\it Synechocystis}, to {\it Cyanothece}
have a circadian clock, which allows them to anticipate the changes
between day and night and to adjust their proteome ahead of time.

To include this into the model, we introduce the notion of the
anticipation time $\Ta$. That is, the cells will compute the target
protein fraction $\phi_\alpha^*(t)$ at time $t$ (see
\erefsrange{JRt}{JAt}) based on the values of
$\nu_\alpha^\beta(t+\Ta)$ at the later time $t+\Ta$. The ribosome
fraction $\chi_\alpha(t)=\phi^*_\alpha(t)$ devoted to making proteins
of sector $\alpha$ at time $t$ is thus determined by the protein
efficiencies $\nu_\alpha^\beta(t+\Ta)$ at a later time $t+\Ta$. This
allows cells to already adjust their proteome before the end of the
day (night) is over, and steer it towards the target protein fractions
set by the efficiencies $\nu_\alpha^\beta(t+T)$ in the following night
(day).

There is one subtlety, which we address in a rather ad-hoc
fashion. The efficiencies $\nuAL(t) = \nuAtL \cN(t) / (\cN(t) + \KA)$
and $\nuCD(t) = \nuCtD \cC(t) / (cC(t) + \KC)$ depend on the
concentrations of cyanophycin and glycogen at time $t$,
respectively. Experiments on plant cells in combination with modeling
\cite{Scialdone:2013bg} suggest that cells might be able to
extrapolate the current concentrations $\cC(t)$ and $\cN(t)$ to
estimate the concentrations at time $t+T$, $\cC(t+T)$ and $\cN(t+T)$,
respectively. While this could be included into our model, we make the
simplication that the cells base the future efficiency based on the
current concentration of the store. 

The target fractions $\phi_\alpha^*(t)$ are thus obtained by solving
\erefsrange{JRt}{JAt} but with the protein efficiencies given by
\begin{align}
\nuC^\beta (t) &\rightarrow L(t+\Ta) \nuCtL + (1-L(t+\Ta)) \nuCtD
\frac{\cC(t)}{\cC(t)+\KC} \elabel{nuCant}\\
\nuA^\beta(t) &\rightarrow L(t+\Ta) \nuAtD \frac{\cN(t)}{\cN(t) +
  \KA} + (1-L(t+\Ta)) \nuAtL \elabel{nuAant},
\end{align}
where, as before, $L(t)$ is an indicator function that is 1 during the
day and 0 during the night.
 
In addition, in this anticipation model, we also take into account
that the protein storing fractions $\phSC$ and $\phSA$ can be made
ahead of time: the synthesis of the glycogen-storing enzymes can
already start before the beginning of the day, while the production of
the cyanophycin-storing enzyme can already start before the beginning
of the night.  As we will see, especially the latter can significantly
enhance the growth rate. Importantly, while the storing enzymes are
synthesized ahead of time, we assume that they become active only when
they need to be, i.e. the cyanophycin-storing enzymes are active only
during the night, while those storing glycogen are only active during
the day.

{\bf Overview full model} The model that takes into the slow proteome
relaxation dynamics but {\em not} anticipation, is given by \eref{ilt}
which gives the instantaneous growth rate $\lambda(t)$, \eref{dphadt}
that describes the evolution of $\phi_\alpha(t)$, and
\erefsrange{chpht}{TargetConstraint}, which are solved to yield
$\chi_\alpha(t) =\phi_\alpha^*(t)$ in \eref{dphadt} and $\sigma(t) =
\sigma^*(t)$ in \eref{ilt}, together with the dynamics for the
concentrations of cyanophycin and glycogen,
\erefstwo{dNdt}{dCdt}. Moreover, in this so-called slow-proteome
  model, we set $\phSC$ to be zero during the night and $\phSA$ to be
  zero during the day, and optimize over the magnitude of their values
  during the day and night, respectively.

  The full model, called the anticipation model, is based on the idea
  that the cell possesses a clock that not only makes it possible to
  anticipate the changes in protein efficiencies $\nu_\alpha^\beta(t)$
  between day and night, but also to express the protein storing
  fractions in an anticipatory fashion. The full model is thus exactly
  the same as the slow-proteome model, except for the following two
  ingredients: 1) the efficiencies $\nuC^\beta(t)$ and $\nuA^\beta(t)$
  in \erefsrange{JRt}{JCt} are replaced by those of
  \erefstwo{nuCant}{nuAant}; 2) the protein-storing fractions
  $\phSC(t)$ and $\phSA(t)$ are optimized not only with respect to
  their magnitude, but also with respect to the timing of their
  expression.

\section{Parameter settings}
In our model, the key parameters that can be varied experimentally are
$\nuCL$, which is determined by the \CO2 and light levels during the
day, $\nuAD$, which is by the \N2 level, and $\nuR$, which can be
varied experimentally via a translational inhibitor such as
chloramphenicol. 
The parameters
$\nuCD$ and $\nuAL$ are set by the nutrient quality of glycogen and
cyanophycin, respectively, while $\nuSC$ and $\nuSA$ are determined by
the efficiencies of the glycogen and cyanophycin storing enzymes,
respectively. We will keep these parameters constant in all the
results that we present below. The parameters are set such that for
the baseline parameter values 
the
average cell-division time is roughly 24h.  The values of $\phRmax$
and $\phRn$ are inspired by those measured for {\it E. coli}
\cite{Hui:2015ig}. The parameters $\phSC$ and $\phSA$ are optimization
parameters, as described above.  We optimize these parameters by
numerically propagating our model for given values of $\phSC$ and
$\phSA$ and numerically computing the average growth rate
$\avg{\lambda}_T=1/T \int_0^T \lambda(t)$ over one period of duration
$T$, which under normal conditions is $T={\rm 24 h}$. 
 
\section{Results}
\subsection{Quasi-equilibrium model}
It is instructive to first consider the scenario in which the
relaxation of the proteome is instantaneous, such that at any moment
in time the protein fractions are optimally balanced based on the
values of the protein efficiencies $\nuC^\beta$ and $\nuA^\beta$ and
the instantanlevels of glycogen, $\cC(t)$, and cyanophycin, $\cN(t)$,
repspectively.

\fref{HeatmapQE}A shows a heat map of the average growth rate over 24h, $\avg{\lambda}_{\rm 24}$ as a function of the fractions of
proteins that store glycogen and cyanophycin, $\phSC$ and $\phSA$,
respectively. The parameters have been set such that the system is
symmetric, $\nuSC=\nuSA$, $\nuCL=\nuAD$, $\KC=\KA$, $c_{\rm G}=c_{\rm
  CP}$, except that the maximum growth rate during the day is slightly
larger than that during the night because $\nuAtL=6/{\rm h}$ is slightly
larger than $\nuCtD=2/{\rm h}$. The prominent feature of the figure is that
even though the system is slightly asymmetric, meaning that the system
could grow during the dark, the storing fractions that maximize the
growth rate are such that the optimal cyanophycin-storing protein
fraction $\phSA^{\rm opt}$ is markedly non-zero, while the optimal
glycogen-storing protein fraction, $\phSC^{\rm opt}$, is essentially
zero.

\begin{figure*}[t]
\centering
\includegraphics[width=2 \columnwidth]{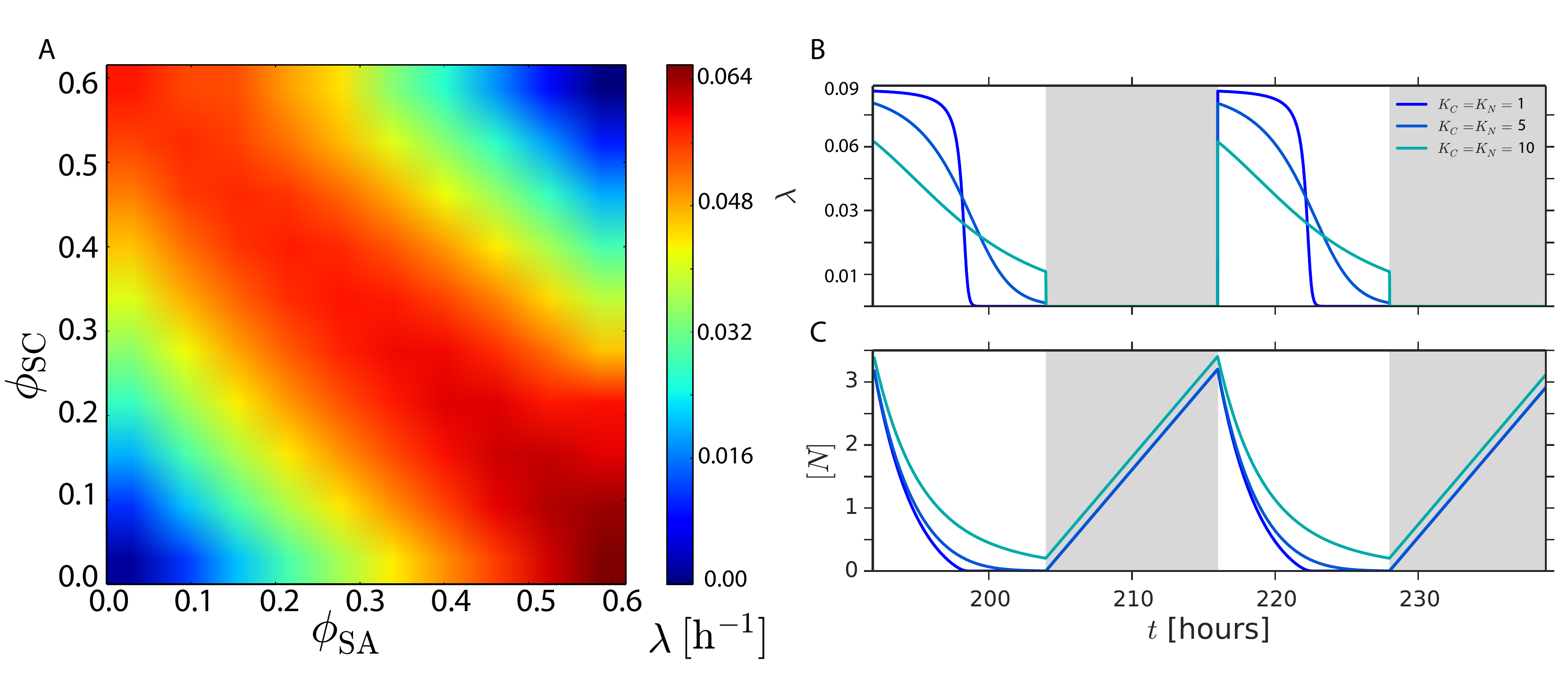}
\caption{Optimal growth strategy in the quasi-equilibrium model. In
  this model, the proteome fractions relax instantaneously and as such
  are always in quasi-equilibrium with the instantaneous levels of
  glycogen, $\cC(t)$, and cyanophycin, $\cN(t)$. In this model, the
  instantaneous growth rate is given by \erefstwo{lambdaL}{lambdaD}
  (or equivalently \erefstwo{lambdaLrw}{lambdaDrw}), while the
  reservoir dynamics is given by \erefstwo{dNdt}{dCdt}. The model is
  nearly symmetric between day and night, $\nuCtL=\nuAtD$,
  $\nuSC=\nuSA$, $c_{\rm G}=c_{\rm CP}$, $\KC=\KA$, except that
  $\nuAL=3/{\rm h}$ is slightly larger than $\nuCD=1/{\rm h}$.  (A)
  Heat map of the average growth rate over 24h, $\avg{\lambda}_{\rm
    24}$, as a function of $\phSC$ and $\phSA$.  The heatmap is
  obtained by numerically propagating \erefstwo{dNdt}{dCdt}, with
  $\lambda(t)$ given by \erefstwo{lambdaL}{lambdaD}, for different
  values of $\phSC$ and $\phSA$; the average growth rate is obtained
  by numerically evaluating $\avg{\lambda}_{\rm 24}=1/24 \int_0^{24}
  \lambda(t)$. It is seen that there exists a combination of storing
  fractions that maximizes the growth rate, $\phSC^{\rm opt}$ and
  $\phSA^{\rm opt}$; moreover, $\phSC^{\rm opt}$ is close to zero,
  while $\phSA^{\rm opt}$ is close to the maximal fraction $\phSANull$
  at which the growth rate becomes zero, see \eref{phSANull}. (B) Time
  traces of $\lambda(t)$ at $\phSC^{\rm opt}$ and $\phSA^{\rm
    opt}$, not only for $\KC=5c_{\rm G}=\KA=5{c}_{\rm CP}$, as in
  panel A, but also for two other values. Clearly, the cells only grow
  during the day. The growth rate during the night is zero, because
  the storing fraction $\phSA^{\rm opt}$ is close to the maximal
  fraction $\phSANull$ at which the growth rate is zero. This shows
  that the the optimal strategy in the quasi-equilibrium model is to
  store as much cyanophycin as possible during the night, because that
  maximizes the growth rate during the day. The explanation of this
  behavior is given in \fref{QEexplain}. (C) Time traces of the
  cyanophycin levels for different values of $\KC=\KA$. Parameter values:
    $\nuCL= 2 / {\rm h}=\nuAD=2 / {\rm h}$; $\nuCtD=2/{\rm h}$;
    $\nuAtL=6 / {\rm h}$; $\nuR=0.2 / {\rm h}$; $\nuSC=\nuSA=0.6 / {\rm h}$; $\KC=5 {\rm
      c}_G=\KA=5{\rm c}_A$. \flabel{HeatmapQE}}
\end{figure*}

To elucidate \fref{HeatmapQE}A, we show in panel B the growth rate
$\lambda(t)$ of the system with the optimal storing fractions
$\phSC^{\rm opt }$ and $\phSA^{\rm opt}$, for three different values
of $\KC=\KA$. Strikingly, the growth rate is zero during the
night. The cells only grow during the day, even though with these
parameters the cells would have the capacity to grow during the night,
had they not to store so much cyanophycin. Indeed, the optimal storing
fraction $\phSA^{\rm opt}$ that maximizes the growth rate is close to
the fraction $\phSANull$ at which the growth rate becomes zero, see
\eref{phSANull}.

The mechanism that underlies the optimal strategy is illustrated in
\fref{QEexplain}. Panel A shows the average growth rate during the day
$\avg{\lambda}_{\rm L}$ as a function of $\phSC$ for different values of
$\phSA$, while panel B of this figure shows the average growth rate
during the night, $\avg{\lambda}_{\rm D}$, as a function of $\phSA$ for
different values of $\phSC$. First of all, note that the maximum
growth rate during the day is only slightly larger than that during
the night---the asymmetry between day and night is indeed (chosen to
be) weak. Yet, the optimal strategy, which maximizes the average
growth rate over 24h, is to not grow at all during
the night. To understand this, note that storing more
cyanophycin during the night will enhance the growth rate during the
day (panel A), yet lower it during the night (panel B). Similarly,
storing more glycogen during the day will raise the growth rate during
the night (right B), yet lower it during the day (panel A). The crux
is that the cost of storing less glycogen during the day---a lower
growth rate at night---decreases when more cyanophycin is stored,
while at the same time the benefit of storing more
cyanophycin---growing faster during the day---is largest when the
amount of stored glycogen is minimal. This tends to favor a strategy
where the maximum amount of cyanophycin is stored during the night,
while a minimal amount of glycogen is stored during the
day. Naturally, the argument also works in the converse direction,
yielding a strategy where the maximal amount of glycogen is stored
during the day and the minimal amount of cyanophycin is stored during
the night. Yet, because the maximal growth rate during the day is
larger than the maximal growth rate during the night, the former
strategy is favored.

\begin{figure*}[t]
\centering
\includegraphics[width=2\columnwidth]{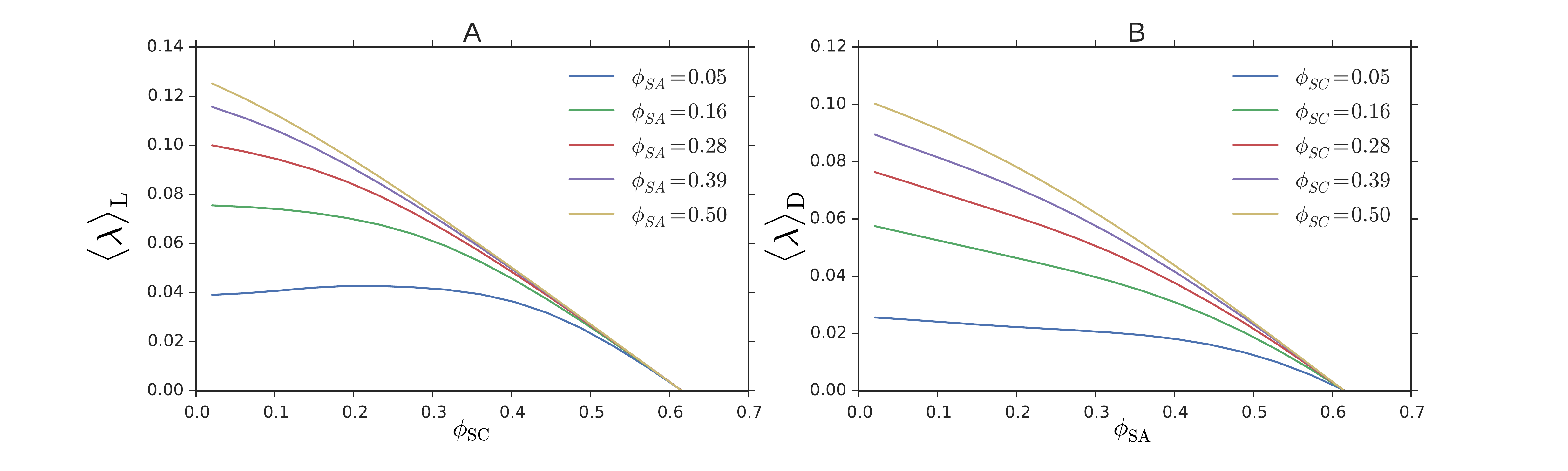}
\caption{Mechanism underlying the optimal strategy that maximizes the
  growth rate in the quasi-equilibrium model, given by
  \erefstwo{lambdaL}{lambdaD} and \erefstwo{dNdt}{dCdt}. (A) The
  average growth rate during the day, $\avg{\lambda}_L$ as a function
  of $\phSC$ for different values of $\phSA$. (B) The average growth
  rate during the night, $\avg{\lambda}_D$ as a function of $\phSA$
  for different values of $\phSC$. These figures have been obtained by
  numerically propagating \erefstwo{lambdaL}{lambdaD} and
  \erefstwo{dNdt}{dCdt} for different combinations of $\phSC$ and
  $\phSA$.  The key point is that the cost of storing less
  glycogen---a lower growth rate at night---decreases when more
  cyanophycin is stored (and vanishes in fact when $\phSA$ approaches
  its maximum $\phSANull$ where the growth rate becomes zero), while
  the benefit of storing more cyanophycin---a higher growth rate
  during the day---increases as less glycogen is stored during the day
  (because $\phSC$ is smaller). This yields an optimal strategy that
  maximizes the growth rate in which the cells exclusively grow during
  the day. Parameter values as in \fref{HeatmapQE}:
   $\nuCL= 2 / {\rm h}=\nuAD=2 / {\rm h}$; $\nuCtD=2/{\rm h}$;
    $\nuAtL=6 / {\rm h}$; $\nuR=0.2 / {\rm h}$; $\nuSC=\nuSA=0.6 / {\rm h}$; $\KC=5 {\rm
      c}_G=\KA=5{\rm c}_A$. Growth rates are in units of $1/{\rm h}$.\flabel{QEexplain}}
\end{figure*}

An important question is how generic this tipping-point strategy in
which cells predominantly grow in one phase of the day, is. What is
essential is that the maximum growth rate during the day,
$\lambda_{\rm max}^{\rm L}$ (\eref{lambdamaxL}), is larger than that
during the day, $\lambda_{\rm max}^{\rm D}$ (\eref{lambdamaxD}). Yet,
the precise values of the efficiencies $\nuC^{\rm L/D}$, $\nuA^{\rm
  L/D}$ tend to be less important, depending on the values of
$\nuSC$ and $\nuSA$. If the model is fully symmetric,
$\nuSC=\nuSA$, $\nuCtD=\nuAtL$, $c_{\rm CP}=c_{\rm G}$, except that
$\nuCL>\nuAD$, then a tipping-point strategy is still favored,
provided that $\nuSC$ and $\nuSA$ are not too large with respect to
$\nuCL$ and $\nuAD$, respectively. The reason is that while increasing
$\nuCL$ with respect to $\nuAD$ increases the maximum growth rate
during the day, which tends to favor growing exlucisely during the
day, it also enhances the capacity to store glycogen (as compared to
that of storing cyanophycin), which tends to favor growing at
night. This effect is particularly pronounced when $\nuSC$ and $\nuSA$
are large compared to $\nuCL$ and $\nuAD$, respectively, because then
the storing rates become limited by $\nuCL$ and $\nuAD$, rather than
being determined by $\nuSC$ and $\nuSA$, respectively (see discussion
below \eref{vmax2}).

The panels of \fref{QEexplain} also reveal that the growth rate is
initially fairly constant, before it markedly drops to zero when the
storing fraction becomes equal to the maximal storing fraction, given
by \eref{phSCNull} for $\phSC$ (panel A) and \eref{phSANull} for
$\phSA$ (panel B). The curves deviate from the linear relationship
between $\lambda$ and the expression of an unused protein, as found in
Scott {\it et al.} \cite{Scott:2010cx}. In fact, also
\erefstwo{lambdaL}{lambdaD} would predict a linear relationship
between $\avg{\lambda}_{{\rm L/D}}$ and $\phSC/\phSA$ if $\nuAL(t)$ and $\nuCD(t)$
were constant in time. However, $\nuAL(t)$ and $\nuCD(t)$ are not
constant in time, because the cyanophycin and glycogen concentrations
decrease with time, as can be seen for the cyanophycin concentration
in panel C of \fref{HeatmapQE} (where $\phSC$ and hence $\cC$ are very
small). As a result of this reservoir depletion, also the growth rate
varies in time.  Moreover, the reservor depletion also underlies the
observation that the rise in the growth rate upon decreasing
$\phSC/\phSA$ becomes less pronounced for low $\phSC/\phSA$
(\fref{QEexplain}): in this regime, the growth rate during the day
(night) is limited by the amount of cyanophycin (glycogen) during the
night (day); decreasing $\phSC$ ($\phSA$) only means that the
reservoir is depleted more rapidly, yielding no significant net
increase in $\avg{\lambda}_{\rm L}$ and $\avg{\lambda}_{\rm D}$; indeed, only by
storing more can the growth rate be enhanced further.

The central prediction of this quasi-equilibrium model is thus that
the cells do not tend to grow at night, as observed experimentally for
{\it Cyanothece} \cite{Cerveny:2009gz}, because that allows it to grow
so much faster during the day that the average growth rate over 24h
increases. However, this quasi-equilibrium model is based on the
assumption that the proteome relaxes instantly, while the relaxation
rate, in the absence of protein degradation, is set by the growth
rate, which, with typical cell-division times of 10-70h \cite{Sinetova:2012bq,Teng:2013cf,Beliaev:2014jf}, is fairly
low for cyanobacteria. In fact, to grow faster, the cell needs to
store more, while the maximum storing capacity is limited by
$\phSCNull$ and $\phSANull$, which depend not only on $\DphRmax$, but
also on $\nuCL$ and $\nuAD$, respectively, as discussed below
\erefstwo{vmax2}{vmax2}. How severe this constraint can be, is
seen in panel B of \fref{HeatmapQE}: for the lowest value of
$\KC=\KA$ shown, the cells grows faster during the beginning of the
day. However, because the cyanophycin stored is then depleted more
radpily (see panel C below), the growth rate drops
sharply well before the end of the
day. Here,  more cyanophycin can not be
stored, simply because $\phSA$ has already reached its maximum,
$\phSANull$. The limited capacity to store thus puts a severe
constraint on the growth rate, which limits the proteome
relaxation rate. Can the cell under these conditions implement the optimal
strategy to maximize the growth rate, as shown in \fref{HeatmapQE} and
\fref{QEexplain}? To address this question, we will turn in the next
section to the influence of the slow proteome dynamics.

\begin{figure*}[t]
\centering
\includegraphics[width=2 \columnwidth]{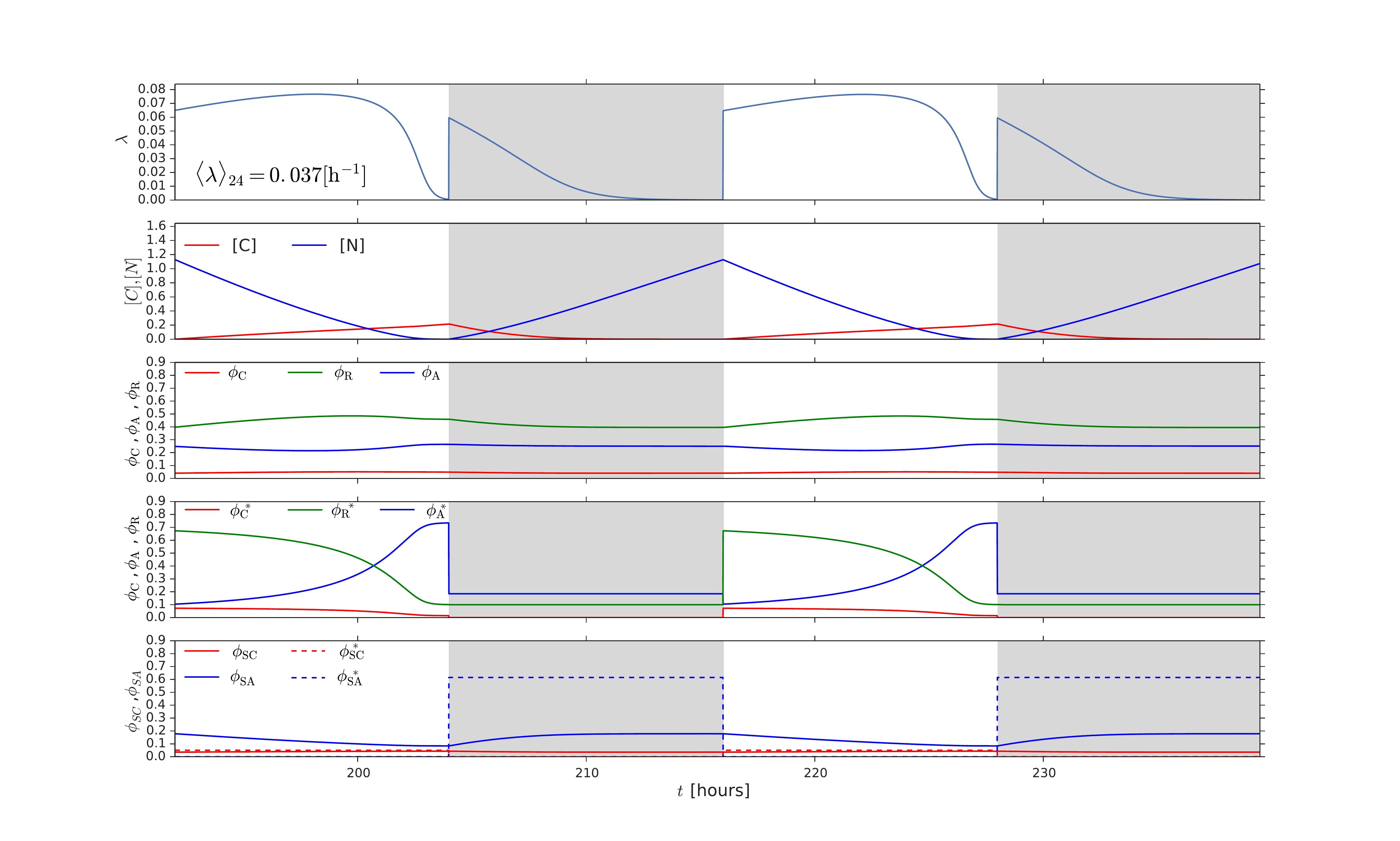}
\caption{Dynamics of the slow-proteome model, given by \eref{ilt} for
  $\lambda(t)$, \eref{dphadt} for $\phi_\alpha(t)$, and
  \erefsrange{chpht}{TargetConstraint}, which are solved to yield
  $\chi_\alpha(t) =\phi_\alpha^*(t)$ in \eref{dphadt} and $\sigma(t) =
  \sigma^*(t)$ in \eref{ilt}, together with \erefstwo{dNdt}{dCdt} for
  the reservoir dynamics. Time traces of the growth rate $\lambda(t)$
  (1/{\rm h}, top row), glycogen levels $\cC(t)$ and cyanophycin
  levels $\cN(t)$ (second row), protein fractions $\phR(t), \phC(t),
  \phA(t)$ (third row), their target fractions $\phR^*(t), \phC^*(t),
  \phA^*(t)$ (fourth row), and the instantaneous storing fractions
  $\phSC(t)$ and $\phSA(t)$ (solid lines, bottom row)), and their
  target fractions $\phSC^*(t)$ and $\phSA^*(t)$ (dashed lines, bottom
  row). Note that because of the slow proteome relaxation resulting
  from the slow growth rate $\lambda(t)$ (see \eref{ilt}), the cell
  also needs to grow significantly during the night in order to
  maintain $\phSA$, necessary to make cyanophycin for growth during
  the day. This is in marked contrast to the dynamics in the
  quasi-equilibrium model, in which the proteome relaxes instantly to
  changing nutrient levels and the cell does not grow at night (see
  \fref{HeatmapQE}). Please also note that the average growth rate is
  significantly lower in this slow-proteome model, $\avg{\lambda}_{\rm
    24}=0.039/{\rm h}$, than in the quasi-equilibrium model,
  $\avg{\lambda}_{\rm 24}=0.064/{\rm h}$. Other parameter values the
  same as in \fref{HeatmapQE}: $\nuCL= 2 / {\rm h}=\nuAD=2 / {\rm h}$;
  $\nuCtD=2/{\rm h}$; $\nuAtL=6 / {\rm h}$; $\nuR=0.2 / {\rm h}$;
  $\nuSC=\nuSA=0.6 / {\rm h}$; $\KC=5 {\rm c}_G=\KA=5{\rm
    c}_A$. \flabel{TimeTracesSP}}
\end{figure*}

\subsection{Slow-proteome model}
\fref{TimeTracesSP} shows time traces of the growth rate, the protein
fractions and the glycogen and cyanophycin levels for our
slow-proteome model. The parameters $\nu_\alpha^\beta$ are identical
to those of the quasi-equilibrium model corresponding to
\fref{HeatmapQE}, yet the magnitudes of $\phSC$ and $\phSA$ have been
optimized to maximize the growth rate $\avg{\lambda}_{\rm 24}$ over
24h (the time windows of $\phSC$ and $\phSA$ expression have not been
optimized in this model, in contrast to in the anticipation model studied
in the next section).

  The first point to note is that the average growth rate in the
  slow-proteome model, $\avg{\lambda}_{\rm 24}=0.037$, is lower
  than in the quasi-equilibrium model, which is
  $\avg{\lambda}_{\rm 24}=0.064$. Clearly, the slow relaxation of
  the proteome drastically lowers the growth rate. The second point is
  that while the cells predominantly grow during the day (top row),
  the growth rate during the night is markedly non-zero near the
  beginning of the night, in marked contrast to the behavior in the quasi-equilibrium model
  (\fref{HeatmapQE}B),

To characterize the growth dynamics further, we show in the second
  row of \fref{TimeTracesSP}
the concentration of cyanophycin and glycogen, respectively. It is
seen that the cyanophycin levels rise during the night, when nitrogen
is stored into cyanophycin, yet fall during the day, when the
cyanophycin provides the nitrogen source for protein production. Near
the end of the day, the cyanophycin levels approach zero,
causing the growth rate to drop to zero. The glycogen levels rise
during the day, which makes it possible to grow during the
night. During the night, however, the glycogen levels rapidly fall,
causing the growth at night to come to a halt.

While the behavior of the reservoir dynamics explains the
time-dependent growth rate $\lambda(t)$ to a large degree, a few
puzzling features remain to be resolved. The first is that the growth
rate at the beginning of the day first rises, even though the levels
of cyanohycin already fall. The second is that the growth rate drops rather
abruptly near the end of the day, even though the concentration of
cyanophycin, $\cN$, is well below the enzyme activation threshold
$\KA$. But perhaps the most important question that needs to be addressed
is why the cells decide to store glycogen and grow at night, given
that the optimal strategy in the quasi-equilibrium model is not to
grow at all during the night (see \fref{HeatmapQE}).

To elucidate these questions, we turn to the time traces of the
protein fractions, shown in the third to fifth row of
\fref{TimeTracesSP}. The third row shows $\phR, \phC, \phA$, while the
fourth row shows the target fractions $\phR^*,\phC^*, \phA^*$ that the
cell aims to reach. The fifth row shows the storing fractions $\phSC$
and $\phSA$, together with their target fractions, $\phSC^*$ and
$\phSA^*$, respectively.

To explain the initial rise of the growth rate, we start by noting
that at the end of the night, $\phA$ is large because the cell needs
to store cyanophycin during the night, which drains nitrogen flux. The
next day, the cell does not need to store nitrogen, while at the
beginning of the day the cyanophycin level---the nitrogen source
during the day---is still high; taken together this means that the
target fraction $\phA^*$ will be relatively low (fourth row). Indeed,
at the beginning of the day, the target fraction $\phA^*$ is smaller
than the current fraction $\phA$, causing $\phA$ to fall
initially. This allows $\phR$ to rise, and since the growth rate is
proportional to $\phR$ (see \eref{ilt}), this tends to raise the growth
rate. The growth thus rises initially, because the proteome slowly
adapts to maximize the growth rate.

\begin{figure*}[t]
\centering
\includegraphics[width=2 \columnwidth]{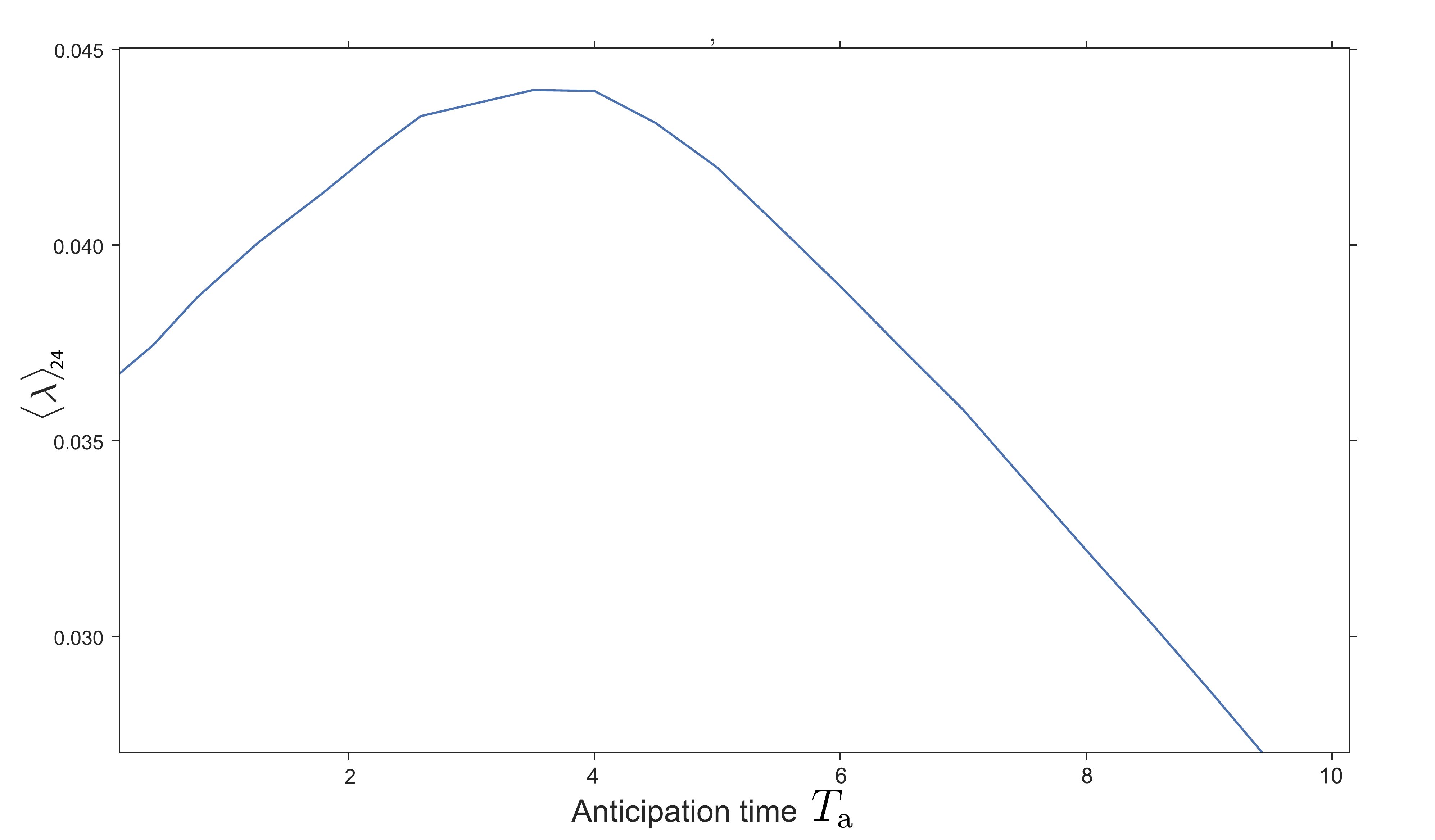}
\caption{The average growth over 24 hours, $\avg{\lambda}_{24}$
  (1/{\rm h}), as a function of the anticipation time $\Ta$ ({\rm h})
  in the full model, including anticipation. The anticipation model is identical to the
  slow-proteome model in that it is given by \eref{ilt} for
  $\lambda(t)$, \eref{dphadt} for $\phi_\alpha(t)$, and
  \erefsrange{chpht}{TargetConstraint}, which are solved to yield
  $\chi_\alpha(t) =\phi_\alpha^*(t)$ in \eref{dphadt} and $\sigma(t) =
  \sigma^*(t)$ in \eref{ilt}, except that the storing fractions
  $\phSC$ and $\phSA$ start to be expressed an anticipation time $\Ta$
  before the beginning of the day and night, respectively; the
  reservoir dynamics is, as for the other models, given by
  \erefstwo{dNdt}{dCdt}. It is seen that there is an optimal
  anticipation time $\Ta^{\rm opt}\approx 4.5 {\rm h}$ that maximizes
  the average growth rate over 24 hours. This maximal growth rate is
  about 20\% higher than in the slow-proteome model (see
  \fref{TimeTracesSP}). The principal reason is that the
  cyanophycins-storing fraction can already be made before the
  beginning of the night, as elucidated in
  \fref{TimeTracesAnticipation}. Other parameter values the same as in
  \fref{HeatmapQE}: $\nuCL= 2 / {\rm h}=\nuAD=2 / {\rm h}$;
    $\nuCtD=2/{\rm h}$; $\nuAtL=6 / {\rm h}$; $\nuR=0.2 / {\rm h}$;
    $\nuSC=\nuSA=0.6 / {\rm h}$; $\KC=5 {\rm c}_G=\KA=5{\rm c}_A$.
  \flabel{GrowthAnticipation}}
\end{figure*}

As time progrresses, the cyanophycin level falls, which causes the
target fraction $\phA^*$ to rise (fourth row). At some point, the
current fraction $\phA$ becomes equal to the target fraction
$\phA^*$. From this moment on, $\phA$ will rise in order to maintain
the flux of nitrogen in the face of the falling cyanophycin
levels. This rise in $\phA$ is accompanied by a drop in $\phC$ and
$\phR$, causing the growth rate to go down.

Finally, why does the cell grow at night?  In this model, the
cyanophycin storing proteins are not made during the day, which means
that then the storing fraction $\phSA$ will fall, because of dilution
due to growth. Inevitably, at the beginning of the night, the fraction
$\phSA$ will always be smaller than that at the end of the night
before. Consequently, $\phSA$ must rise to move towards the target
fraction $\phSA^*$, which in this case is close to the maximium at
which the growth rate is zero, $\phSANull$ (dashed blue line in last
row). However, in the absence of protein degradation, the proteome can
only relax because of growth, and, indeed, this is the reason why the
cell needs to grow during the night: without growth, $\phSA$ would
eventually become zero, and no cyanophycin could be stored. During the
night, new storing proteins have to be made, in order to compensate
for the drop in $\phSA$ resulting from dilution during the day.

Lastly, in order to grow during the night, the cell needs to store
glycogen during the day, which explains why $\phSC$ is non-zero during
the day. The cell thus adopts a mixed strategy in which it grows
during the day and during the night, because this is the optimal
strategy in the presence of slow proteome relaxation. In the next
section, we will study whether anticipation makes it possible to
counteract the detrimental effects of slow proteome relaxation, by
initiating a response ahead of time.

\section{Anticipation}
To study the importance of anticipation, we first consider the
scenario where the cell can express the storing fractions $\phSC$ and
$\phSA$ before the beginning of the day and the night, respectively;
here, we thus do not consider the possibility that cells can
anticipate the changes in the protein efficiencies $\nu_\alpha^\beta$
(see \erefstwo{nuCant}{nuAant}). More specifically, we consider 4
optimization parameters: the magnitudes of $\phSC$ and $\phSA$ and the
timings of their expression; to limit the optimisation space, we take
the duration of the expression window to be constant, namely
12h. Performing the optimisation, we observed that the growth-rate
dependence on the expression timing of $\phSC$ was rather weak,
because, as we will see below, the optimal $\phSC$ is very small. We
therefore considered one anticipation time $\Ta$, which determines the
times $k 24 -\Ta$ and $12 + k 24 - \Ta$, with $k=0,1,2,\dots$, from
which $\phSC$ and $\phSA$ respectively are expressed for 12 hours at
constant values, respectively.  This limits the optimisation space to
3 parameters: the magnitudes of $\phSA$ and $\phSC$, respectively, and
the anticipation time $\Ta$.

To analyze the importance of anticipation, we optimized the growth
rate over $\phSA$ and $\phSC$ for each value of $\Ta$, for the same
set of parameters as in
\frefsrange{HeatmapQE}{TimeTracesSP}. \fref{GrowthAnticipation} shows
the result. It is seen that expressing the storing enzymes about 4.5
hours before the beginning of the next part of the day can speed up
growth by about 20\%.

\begin{figure*}[t]
\centering
\includegraphics[width=2 \columnwidth]{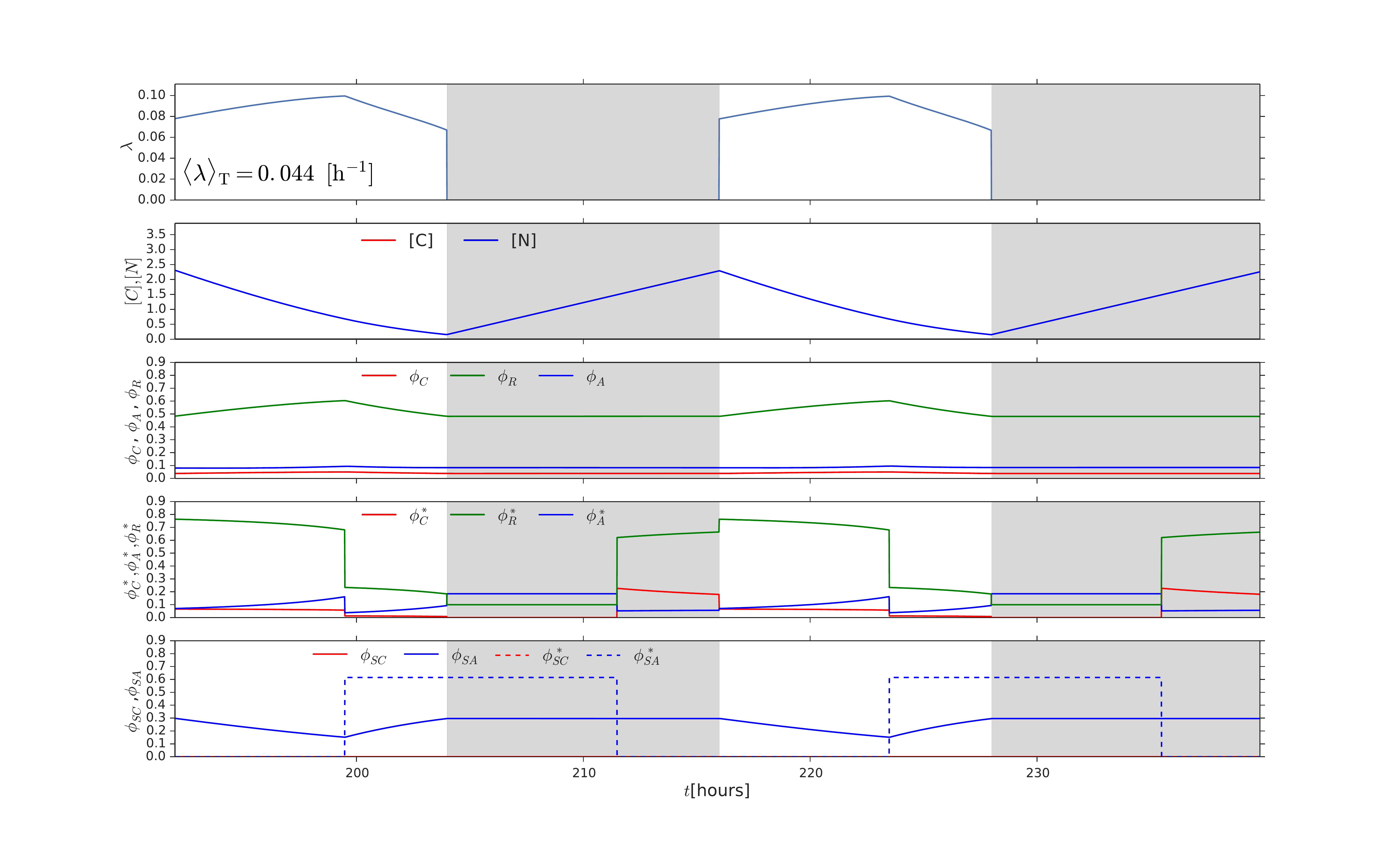}
\caption{Dynamics of the full model, including anticipation. The model
  is identical to that of \fref{GrowthAnticipation} with the
  anticipation time $T_a$ equal to its optimal value $\Ta^{\rm opt}=4.5{\rm h}$. Time traces of the growth rate $\lambda(t)$
  (1/{\rm h}, top row), glycogen levels $\cC(t)$ and cyanophycin
  levels $\cN(t)$ (second row), protein fractions $\phR(t), \phC(t),
  \phA(t)$ (third row), their target fractions $\phR^*(t), \phC^*(t),
  \phA^*(t)$ (fourth row), and the instantaneous storing fractions
  $\phSC(t)$ and $\phSA(t)$ (solid lines, bottom row)), and their
  target fractions $\phSC^*(t)$ and $\phSA^*(t)$ (dashed lines, bottom
  row). The average growth rate over 24 hours in this model,
  $\avg{\lambda}_{24}=0.044/{\rm h}$, is about 20\% higher than in the
  slow proteome model. Note also that the growth rate during the day
  first rises because the proteome is still adapting to the nutrient
  levels (top row); however, $\Ta=4.5$h before the beginning of the
  night, the growth rate goes down, because the cell prepares for the
  night by expressing the cyanohycin storing fraction $\phSA$ (bottom
  panel). Before the end of the day, $\phSA$ has reached a level that
  is sufficient to store enough cyanophycin during the night for
  fueling growth the next day. Concomitantly, the growth rate is now
  zero during the night, in contrast to the scenario in the
  slow-proteome model where $\phSA$ has to be made during the night,
  and the cells therefore have to grow during the night
  \fref{TimeTracesSP}. Other parameter values the same as in
  \fref{HeatmapQE}: $\nuCL= 2 / {\rm h}=\nuAD=2 / {\rm h}$;
    $\nuCtD=2/{\rm h}$; $\nuAtL=6 / {\rm h}$; $\nuR=0.2 / {\rm h}$;
    $\nuSC=\nuSA=0.6 / {\rm h}$; $\KC=5 {\rm c}_G=\KA=5{\rm
      c}_A$.\flabel{TimeTracesAnticipation}}
\end{figure*}

To elucidate this behavior, we show in \fref{TimeTracesAnticipation}
the time traces for the optimal anticipation time $\Ta^{\rm opt}=4.5
{\rm h}$ that maximizes the growth rate
(\fref{GrowthAnticipation}). The top row shows that, as in the
slow-proteome model, the growth rate first rises at the beginning of
the day, because the proteome still adapts to the changing nutrient
levels. However, at about $\Ta^{\rm opt}=4.5{\rm h}$ before the end of
the day, the growth rate goes down markedly. This is because the cell
starts to express the proteins $\phSA$ that store cyanophycin during
the night (bottom row). Clearly, there is a cost to anticipation: it
lowers the instantaneous growth rate. The cell should therefore not
express the cyanophycin-storing proteins too early in the day. Yet,
expressing cyanophycin-storing enzymes already during the day also has
a marked benefit: it makes it possible to reach a sufficiently high
level of $\phSA$ before the beginning of the night such that enough
cyanophycin can be stored during the night. The cell therefore does
not need to grow during the night to raise $\phSA$, as in the
slow-proteome model; indeed, even though $\phSA$ does not rise during
the night, the level is much higher than the average level in the
slow-proteome model, so that more cyanophycin is stored during the
night, as a result of which the cells grow much faster during the
day (compare with \fref{TimeTracesSP}). Anticipation thus makes it
possible to implement the optimal growth strategy as revealed by the
quasi-equilibrium model (see \fref{HeatmapQE}), which is to grow
exclusively during the day, as observed experimentally.

We also considered anticipation of $\nuC^\beta$ and $\nuA^\beta$, as
described around \erefstwo{nuCant}{nuAant}. However, because the
growth rate is zero at night, the benefit of optimising $\nuC^\beta,
\nuA^\beta$ is marginal, for two reasons.  Firstly, because the cell
cannot grow at night, it cannot adjust the proteome before the
beginning of the day. Secondly, adjusting the proteome fractions
during the day based on the anticipated efficienies $\nuCD$ and
$\nuAD$ during the night would lower the instantaneous growth rate,
because the instantaneous protein fractions $\phi_\alpha$ would become
suboptimal, i.e. not given by the current efficiencies $\nuCL$ and
$\nuAL$.


\section{Discussion}
The power of the framework of Hwa and coworkers is that it provides a
coarse-grained description of the proteome with only a limited number
of sectors, characterized by enzyme efficiencies that can be measured
experimentally
\cite{Scott:2010cx,You:2013ey,Hui:2015ig,Erickson:2017ic,Mori:2017cz}. We
therefore sought to develop a minimal model, consisting of a small
number of sectors that can be characterized experimentally, also given
the fact that as yet there is no experimental data that warrants a
more detailed model. Nonetheless, even though the model consists of
only 3 main sectors and 2 storing sectors, the dynamical behavior of
our model is already very rich. Specifically, our analysis shows that
the requirement to store carbon and nitrogen means that the cells tend
to adopt an extreme strategy in which they exclusively grow during the
day.  The fundamental reason is contained in the growth laws uncovered
in
refs. \cite{Scott:2010cx,You:2013ey,Hui:2015ig,Erickson:2017ic,Mori:2017cz}:
storing more glycogen during the day will increase the growth rate
during the night, yet this benefit decreases as more
cyanophycin-storing enzymes are expressed during the night (and
vanishes in fact when this fraction approaches its maximum at which
the growth rate becomes zero, see 
\fref{QEexplain}); at the same time, the benefit of
storing more cyanophycin during the night---growing faster during the
day---increases as less glycogen is stored during the day. The interplay between these two effects creates a
positive feedback loop in which the cells store as much cyanophycin as
possible during the night and as little glycogen as needed during the
day to maximize the growth rate during the day.

\begin{figure*}[t]
\centering
\includegraphics[width=2 \columnwidth]{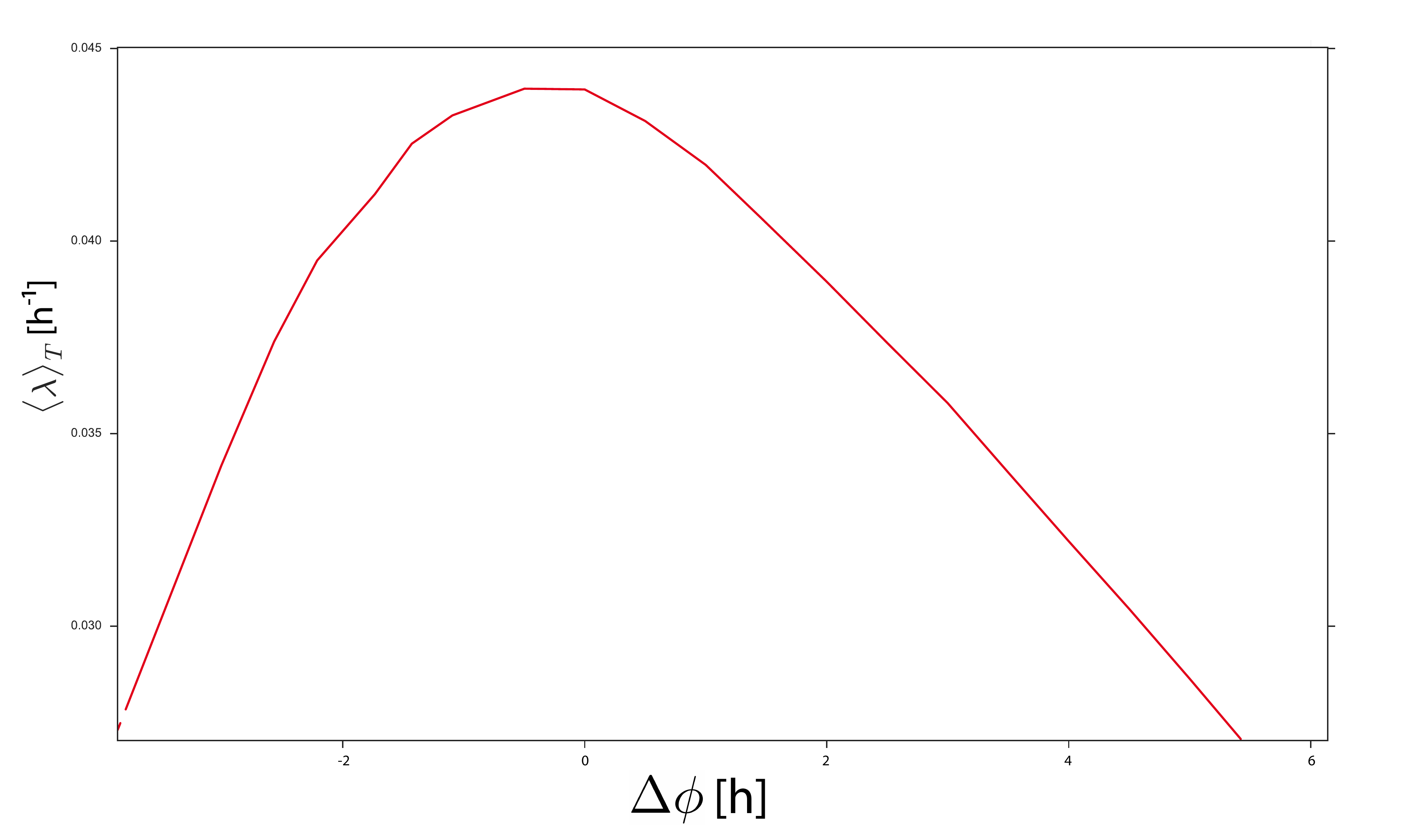}
\caption{Changing the phase of the clock from its optimal value can
  significantly reduce the growth rate. The model is identical to that
  of \fref{TimeTracesAnticipation}, meaning that a phase shift $\Delta
  \phi = 0$ corresponds to the full model with the optimal
  anticipation time $T_a^{\rm opt}=4.5{\rm h}$. When $\Delta \phi$ is
  changed away from 0, all parameters, including the magnitudes of
  $\phSC$ and $\phSA$, are kept constant, except for the time windows in
  which $\phSC$ and $\phSA$ are expressed: these windows are shifted
  by an amount $\Delta \phi$. Indeed, while in
  \fref{GrowthAnticipation} the magnitudes of $\phSC$ and $\phSA$ are
  optimized for each value of $T_a$, here the values of $\phSC$ and
  $\phSA$ remain equal to those corresponding to $T_a^{\rm opt}$. Other
  parameter values the same as in \fref{HeatmapQE}: $\nuCL= 2 / {\rm
    h}=\nuAD=2 / {\rm h}$; $\nuCtD=2/{\rm h}$; $\nuAtL=6 / {\rm h}$;
  $\nuR=0.2 / {\rm h}$; $\nuSC=\nuSA=0.6 / {\rm h}$; $\KC=5 {\rm
    c}_G=\KA=5{\rm c}_A$.\flabel{PhaseShift}}
\end{figure*}

Our analysis also reveals that the slow relaxation of the proteome
creates a severe challenge in implementing the optimal strategy. In
the absence of protein degradation, the cells need to grow in order to
adjust their proteome. Yet, cyanobacterial cells grow slowly, which
means that the relaxation time will be long compared to the 24h period
of the day-night rhythm. In fact, to reach the required
cyanophycin-storing fraction, cells would need to grow significantly
during the night, in the absence of anticipation. Indeed, the
principal benefit of having a circadian clock, according to our model,
is that by knowing the time the cyanobacterium can anticipate the
shift from day to night and express the storing proteins ahead of
time. Interestingly, this prediction appears to be supported by recent
mass-spectrometry proteomics data and RNA-sequencing transcriptomics
data on {\it Cyanothece}: the expression of cyanophycin is highest in
the late {\em light} and progressively dimishes during the night into
the early light \cite{Welkie:2014ho}.

In a beautiful series of experiments, Johnson and coworkers showed
that circadian clocks can provide a fitness benefit to organisms that
live in a rhythmic, circadian environment
\cite{Woelfle:2004cq,Ouyang:1998wp}. Mutants of {\it S. elongatus}
with different intrinsic clock periods were competed with wild-type
strains and with each other, and the strain whose intrinsic clock
period most closely matched that of the light-dark (LD) cycle won the
competition \cite{Woelfle:2004cq,Ouyang:1998wp}. When the intrinsic
period of the clock and the period of the LD cycle are altered
with respect to one another, then the period of the (driven) clock
remains equal to that of the driving signal, as long as the clock remains
phase locked to the LD cycle \cite{Monti:2018hs}. However,
their phase relationship will change \cite{Monti:2018hs}. This altered
phase relationship is probably the reason why strains with
`non-resonant' clock rhythms have a lower growth rate
\cite{Ouyang:1998wp}. We investigated whether, according to our model,
{\it Cyanothece} would exhibit similar behavior. To this end, we
started from the idea that changing the intrinsic clock period keeping
the period of the LD cycle equal to 24h, will alter the phase
of the clock. We thus computed the average growth rate
$\avg{\lambda}_{24}$ as a function of the phase
shift $\Delta \phi$. \fref{PhaseShift} shows the result. It is seen that changing
the phase of the clock from its optimal value can reduce
the growth rate by more than 10\%. With an incorrectly set clock the cells can
no longer accurately anticipate dawn and concomitantly start the production of the
cyanophycin storing enzymes either too late or too early.

In our model, cyanophycin serves exlucisely as a source of nitrogen,
which is a reasonable starting point given that cyanophycin is very
rich in nitrogen. However, cyanophycin also contains carbon and it has
indeed been speculated that it also provides a carbon store
\cite{Aryal:2011bc}. Our model could be extended to include
this. While the benefit of providing a carbon source during the day
might be small in the presence of high \CO2 and light levels, the cost
of draining carbon flux at night might be more significant---this
effect could be included by adding a term to the equation for the
carbon flux (\eref{JC}), representing the carbon flux into cyanophycin
during the night. Including this effect in the model
will also raise the required levels of glycogen.

While the dynamics of our minimal model is already complex, it seems
natural to increase the number of sectors as more data becomes
available. In particular, it might be of interest to distinguish
between proteins of a given sector that are generic, i.e. expressed at
signifincant levels both during the day and during the night, and
proteins that are specific to one part of the day, such as the
photosynthesis components. The challenge will be to define major
subsectors and devise experiments which make it possible to measure
the associated enzyme efficiences.

Another natural extension of our model is to include protein
degradation. First of all, active protein degradation makes it
possible to increase the proteome relaxation rate. While active
protein degradation by itself tends to slow down the growth rate,
reaching the optimal proteome partitioning faster might offset this
cost. Secondly, some proteins tend to be unstable, meaning that
degradation by spontaneous decay is inevitable. In our full model, the
amount of glycogen stored is vanishingly small, because our model only
considers glycogen as a source of carbon for protein synthesis and the
cells do not grow during the night. At the same time, it is well known
that cyanobacteria store glycogen. Some of the stored glycogen will be
essential for providing the energy to run maintenance processes, such
as DNA repair, or to drive the cyanophycin-storing reactions---storing
cyanophycin is ATP consuming \cite{Aboulmagd:2001cy}. However, it is
also possible that glycogen is needed to synthesize those proteins
that have decayed significantly during the night, such as the
components of the protein synthesis machinery. It would then be
interesting to see whether including this into the model would yield
the prediction that it is beneficial to start expressing these
proteins before the end of the night, as observed experimentally
\cite{Welkie:2014ho}.

We have focused here on the cyanobacterium {\it Cyanothece}. However,
the application of our framework to cyanobacteria such as {\it
  Synechococcus} and {\it Synechocystis} predicts that also these
bacteria tend to grow predominantly during the day (data not shown),
as observed experimentally \cite{Mori:1996ws}. If the marginal cost of
storing glycogen---the reduction in the growth rate during the
day---is higher than the marginal benefit---the increase in the growth
rate during the night---then the optimal strategy is to not store any
glycogen at all for growth during the night, and hence exclusively
grow during the day. While this observation may explain why these
cyanobacteria predominantly grow during the day, it does not explain
why these bacteria have a clock. Indeed, the mechanism by
which a clock provides a benefit to {\it Cyanothece} as predicted by
our model---namely that it allows the cell to make storing proteins
before it stops growing---does not apply to {\it Synechococcus} and
{\it Synechocystis}. It is tempting to speculate that the latter
cyanobacteria possess a clock because that enables them to replace the
photosynthesis proteins which have decayed during the night before
the sun rises again.



\end{document}